\documentclass{mnras}

\usepackage{natbib}
\usepackage{graphicx}
\usepackage{epstopdf}
\usepackage[T1]{fontenc}
\usepackage{aecompl}
\usepackage{mnras-bst}
\usepackage{hyperref}
\bibpunct{(}{)}{;}{a}{}{,}

\usepackage{amsmath}

\newcommand\omicron{o}
\newcommand\ksi{\xi}

\title[Polarization of FGK dwarfs]{The intrinsic and interstellar broadband linear polarization of nearby FGK dwarfs}
\author[D.V.~Cotton et al.]{Daniel V. Cotton$^{1,2}$\thanks{E-mail: d.cotton@unsw.edu.au}, 
Jonathan P. Marshall$^{1,2,3}$, Jeremy Bailey$^{1,2}$,
\newauthor Lucyna Kedziora-Chudczer$^{1,2}$, Kimberly Bott$^{1,2,4,5}$, Stephen C. Marsden$^{3}$,
\newauthor and Bradley D. Carter$^{3}$ \\
$^{1}$School of Physics, UNSW Australia, NSW 2052, Australia.\\
$^{2}$Australian Centre for Astrobiology, UNSW Australia, NSW 2052, Australia. \\
$^{3}$Computational Engineering and Science Research Centre, University of Southern Queensland, Toowoomba, Qld. 4350, Australia. \\
$^{4}$University of Washington Astronomy Department, Box 351580, UW Seattle, WA 98195, USA. \\
$^{5}$NASA Astrobiology Institute Virtual Planetary Laboratory, Box 351580, UW Seattle, WA 98195, USA.}
\begin{document}

\date{Accepted . Received ; in original form }

\pagerange{\pageref{firstpage}--\pageref{lastpage}} \pubyear{2016}

\maketitle

\label{firstpage}

\begin{abstract}We present linear polarization measurements of nearby FGK dwarfs to parts-per-million (ppm) precision. Before making any allowance for interstellar polarization, we found that the active stars within the sample have a mean polarization of 28.5~$\pm$~2.2 ppm while the inactive stars have a mean of 9.6~$\pm$~1.5 ppm. Amongst inactive stars we initially found no difference between debris disk host stars (9.1~$\pm$~2.5 ppm) and the other FGK dwarfs (9.9~$\pm$~1.9 ppm). We develop a model for the magnitude and direction of interstellar polarization for nearby stars. When we correct the observations for the estimated interstellar polarization we obtain 23.0~$\pm$~2.2 ppm for the active stars, 7.8~$\pm$~2.9 ppm for the inactive debris disk host stars and 2.9~$\pm$~1.9 ppm for the other inactive stars. The data indicates that whilst some debris disk host stars are intrinsically polarized most inactive FGK dwarfs have negligible intrinsic polarization, but that active dwarfs have intrinsic polarization at levels ranging up to $\sim$~45 ppm. We briefly consider a number of mechanisms, and suggest differential saturation of spectral lines in the presence of magnetic fields is the best able to explain the polarization seen in active dwarfs. The results have implications for current attempts to detect polarized reflected light from hot Jupiters by looking at the combined light of the star and planet. \end{abstract}

\section{Introduction}

Scattering from cloud particles in planetary atmospheres polarizes light. A number of efforts have been made \citep{lucas09,wiktorowicz09,berdyugina08,berdyugina11} and are underway \citep{wiktorowicz15b,bott16} to detect reflected light from hot Jupiter atmospheres in the combined light of the star and planet using broadband aperture polarimetry. A signal should show up as a variable polarization around the orbital cycle, with a peak near $\sim$~20 ppm in blue light expected in the most promising systems \citep{seager00,bott16}. In such work it is usually assumed that the light from the star is unpolarized but there is very little evidence to support such an assertion at the needed precision.

High precision polarimetric surveys of nearby stars have been conducted by \citet{bailey10} in a red (575-1025 nm) bandpass, and \citet{cotton16,cotton16b} in the SDSS $g^{\rm \prime}$ band (green) -- which is more relevant to exoplanet work. These surveys identified intrinsic polarization from extreme stellar types (late giants, B- and Be-stars, Ap stars) and some debris disk systems, but none from ordinary main sequence stars. However, both of these surveys were magnitude limited, and as a result included very few later type main sequence stars. The aim of the present study is to extend that work further down the main sequence.

Parts-per-million polarimetry of the fainter main sequence stars has only recently become possible \citep{hough06,kochukhov11}, and to date there are no convincing determinations of the level of broadband polarization in FGK dwarfs. \citet{kemp87} used a special instrumental arrangement to measure the whole disc of the quiet Sun, obtaining a linear polarization of $<$~0.3 ppm. Yet, there is some reason to suspect that broadband polarization may manifest in FGK dwarfs. The (transverse component of) magnetic fields associated with starspot regions on the Sun produce linear polarization in spectral lines as a result of the Zeeman Effect \citep{donati09}. Where there are sufficient spectral lines blanketing a band, the combined effect may be enough to produce a measurable linear polarization; the mechanism is properly known as \textit{differential saturation}\footnote{Many of the authors we cite on this topic refer to the phenomenon as magnetic intensification rather than differential saturation. However magnetic intensification \citep{babcock49} does not necessarily involve the line-blanketing necessary to generate broadband linear polarization, and so we prefer differential saturation here.} after the differential saturation of the $\pi$ and $\sigma$ components of the Zeeman multiplet that occurs in the transfer of radiation in a stellar atmosphere \citep{bagnulo95}. Early on \citet{tinbergen81} invoked this mechanism in what they described as an ``attractive hypothesis'' to explain a weak trend to higher polarizations with later spectral type in stars from F0 onwards. The idea was developed by many including \citet{landideglinnocenti82,leroy90,huovelin91,saar93,stift97}, who made calculations based on fields localised in starspots. In contrast to those predictions, more recent spectropolarimetric measurements of circular polarization have revealed large-scale magnetic fields of varying complexity, that appear not to be associated with cool spots \citep{donati09,jeffers14,morgenthaler12,fares10}. Linear polarization has been detected in the spectral lines of active cool stars \citep{kochukhov11,rosen13,rosen15}, which can, in principle, be used to derive the broadband linear polarization \citep{wade00}. Yet, to date, there are no satisfying systematic measurements of the effect of such magnetic fields in linear broadband polarization. \citet{huovelin88} made measurements that attempted to correlate broadband linear polarization with the activity indicator $log(R'_{HK})$. However, the sensitivity of their instrument meant they had to rely on statistical techniques that only considered measurements 2-sigma from the mean. According to \citet{clarke91} these observations were contested at the time by \citet{leroy89} and others as being unreliable due to problems with scattered moonlight (particularly in U-band), and he would later describe this area of research as ``abandoned'' \citep{clarke10}. Yet, \citet{alekseev03} and most recently \citet{patel16} have copied the multi-band approach of \citet{huovelin88}, with similar results -- finding increased levels of polarization in shorter wavelength bands for active dwarfs, which \citet{patel16} assign to a combination of differential saturation and scattering processes.

Aside from possibly differential saturation, in FGK dwarfs measurable polarization will be present in some debris disk systems, such as those observed by \citet{hough06,bailey10,wiktorowicz08}, due to scattering from the dust grains in the disks. The magnitude and direction of the polarization is a function of the disk geometry with respect to the aperture and line of sight, and the size, shape, composition and porosity of the dust grains. \citet{clarke10}'s comprehensive text book relates no other detections or prospects for detection in solar type stars. The sole exception being the young ($\sim$~70~Myr) star HD 129333 studied by \citet{elias90} which exhibited an unexplained polarization angle variation unconnected to its rotational period. In this case the authors suggested that the most likely cause of the polarization was scattering from a circumstellar envelope modulated by the motion of an unseen companion.

In contrast to the \textit{intrinsic polarization} related to the stars themselves (or their immediate surrounds), the light reaching us from all stars is extrinsically polarized by aligned dust grains in the interstellar medium (ISM). This \textit{interstellar polarization} is largely constant for any given star system, but acts to confound measurements of intrinsic polarization. In distant stars the sheer magnitude of interstellar polarization can swamp any intrinsic signal. In nearby space the region known as the Local Hot Bubble (LHB), extending out to $\sim$~75 to 150 pc from the Sun, is largely devoid of dust and gas. In this region the ISM is polarized at a rate of $\sim$~0.2 to 2 ppm/pc \citep{cotton16}. This is small, at least an order of magnitude smaller than the region beyond the LHB \citep{behr59}, but when seeking intrinsic effects at the level of tens of ppm, it is significant, and needs to be subtracted. Frisch et al. (submitted) are working on improving their mapping of the ISM field in nearby space. Broadband stellar optical polarimetry will help in this task \citep{frisch12}, but at present the data within 50 pc are sparse (especially at southern latitudes). As a result the local interstellar polarization tends to be neglected, as it has been in the studies of active late dwarfs mentioned above.

In the following sections of this paper we describe a polarimetric survey of FGK dwarfs. We begin with details of the observations (Section \ref{sec:obs}) and the results of those observations (Section \ref{sec:results}). We then make an initial analysis of the results to attempt to identify statistical differences between active stars and inactive stars, and between debris disk host stars and ordinary FGK dwarfs (Section \ref{sec:stats}). After that we add the appropriate parts of our data set to measurements from the literature in order to develop a model to describe interstellar polarization in the nearby ISM (Section \ref{sec:interstellar}); some comments are made about the nature of the ISM in passing. In Section \ref{sec:isub} we carry out a vector subtraction using our simple model to calculate the intrinsic polarization of the programme stars. Following this we examine and discuss the intrinsic polarization in ordinary FGK dwarfs (Section \ref{sec:FGK}), debris disk host systems (Section \ref{sec:dd}) and active stars (Section \ref{sec:active}). Our conclusions are presented in Section \ref{sec:conclusions}.

\section{Observations}
\label{sec:obs}

\subsection{The sample stars}

Our aim here was to investigate intrinsic polarization toward the end of the main sequence; specifically F, G and K types of which there are few examples in our previous surveys \citep{bailey10,cotton16,cotton16b}. To do this effectively we aimed for a polarimetric precision of less than 10 ppm per target. To achieve this in a time-efficient manner we imposed a magnitude limit of 6.0 in selecting the programme stars. The mean precision finally achieved was 6.9 ppm, with the worst for any target being 10.1 ppm.

\begin{table*}
\caption{Properties of survey stars.}
\tabcolsep 2.5 pt
\centering
\begin{tabular}{lllrrlrrccccrrrl}
\hline
HD      & HIP       & Other Names           & V         & $B-V$     & Spectral          & Sep$^a$   & Dist    & RA            &  Dec          & \multicolumn{2}{c}{Galactic}  & Notes$^b$ \\
        &           &                       & mag       & mag       & Type              & ($\arcsec$) & (pc)      &(hh mm ss.s)  & (dd mm ss)  & $l$ ($\degr$) & $b$ ($\degr$)   &          \\
\hline
\multicolumn{13}{l}{\textit{Ordinary FGK Dwarfs}} \\                
10360	&   7751 B  &	p Eri A	            &	5.96	&	0.88	&	K2V             & 11.22     &	 7.8	& 01 39	47.6	& -56 11 36	&	289.59	&	-59.67	        &   \\
23754	&	17651	&	$\tau^6$ Eri	    &	4.20	&	0.45	&	F5IV-V	        &           &	17.6	& 03 46	50.9	& -23 14 59	&	217.35	&	-50.33	        &   \\
30652	&	22449	&	$\pi^3$ Ori 	    &	3.19	&	0.44	&	F6V	            &           &	 8.1	& 04 49	50.4	&  06 57 41	&	191.45	&	-23.07	        &   Var     \\
38393	&	27072	&	$\gamma$ Lep        &	3.60	&	0.47	&	F6V	            &           &	 8.9	& 05 44	27.8	& -22 26 54	&	226.80	&	-24.27	        &   \\
64096	&	38382	&	9 Pup               &	5.16	&	0.60	&	G0V$^c$	        &  0.62     &	16.5	& 07 51	46.3	& -13 53 53	&	232.27	&	  6.62	        &   \\
102365	&	57443	&	HR 4523, 66 G Cen	&	4.88	&	0.67	&	G2V+M4V         & 22.99     &	 9.2	& 11 46	31.1	& -40 30 01	&	289.80	&	 20.71	        &   EP      \\
102870	&	57757	&	$\beta$ Vir	        &	3.60	&	0.55	&	F9V	            &           &	10.9	& 11 50	41.7	&  01 45 53	&	270.52	&	 60.75	        &   LP      \\
114613	&	64408	&	GJ 501.2	        &	4.85	&	0.70	&	G3V	            &           &	20.7	& 13 12	03.2	& -37 48 11	&	307.42	&	 24.89	        &   EP      \\
119756	&	67153	&	i Cen	            &	4.23	&	0.38	&	F2V	            &< 0.01$^d$ &	19.4	& 13 45	41.2	& -33 02 37	&	315.85	&	 28.47	        &   \\
132052	&	73165	&	16 Lib	            &	4.49	&	0.32	&	F2V	            &           &	26.9	& 14 57	11.0	& -04 20 47	&	351.73	&	 46.27	        &   \\
141004	&	77257	&	$\lambda$ Ser       &	4.42	&	0.61	&	G0IV-V	        & ($^e$)    &	12.1	& 15 46	26.6	&  07 21 11	&	 15.69	&	 44.10	        &   \\
156384	&	84709	&	GJ 667	            &	5.89	&	1.04    &   K3V+K5V$^f$     &  1.82     &    6.8	& 17 18	57.2	& -34 59 23	&	351.84	&	  1.42	        &   \\
197692	&	102485	&	$\psi$ Cap	        &	4.15	&	0.39	&	F5V	            &           &	14.7	& 20 46	05.7	& -25 16 15	&	 20.00	&	-35.50	        &   \\
209100	&	108870	&	$\epsilon$ Ind	    &	4.69	&	1.06	&	K5V$^g$	            &           &	 3.6	& 22 03	21.7	& -56 47 10	&	336.19	&	-48.04	        &   LP, PMS$^h$ \\
\hline
\multicolumn{13}{l}{\textit{Debris Disk Host Stars}} \\
1581	&   1599	&	$\zeta$ Tuc	        &	4.23	&	0.57	&	F9.5V	        &           &	 8.6	& 00 20	04.3	& -64 52 29	&	308.32	&	-51.93	        &   \\
10700	&	8102	&	$\tau$ Cet	        &	3.50	&	0.72	&	G8.5V	        &           &	 3.7	& 01 44	04.1	& -15 56 15	&	173.11	&	-73.44	        &   HD      \\
20794	&	15510	&	e Eri	            &	4.27	&	0.71	&	G8V	            &           &	 6.0	& 03 19	55.7	& -43 04 11	&	250.75	&	-56.08	        &   HD, EP  \\
20807	&	15371	&	$\zeta^2$ Ret	    &	5.24	&	0.60	&	G0V	            &           &	12.0	& 03 18	12.8	& -62 30 23	&	278.98	&	-47.22	        &   \\
105211  &   59072   &   $\eta$ Cru          &   4.15    &   0.32    &   F2V$^i$         & 48.41     &   19.8    & 12 06 52.9    & -64 36 49 &   298.18  &    -2.15          &   \\
109085	&	61174	&	$\eta$ Crv	        &	4.31	&	0.38	&	F2V             &           &	18.3	& 12 32	04.2	& -16 11 46	&	296.18	&	 46.42	        &   Var     \\
115617	&	64924	&	61 Vir	            &	4.74	&	0.70	&	G7V	            &           &	 8.6	& 13 18	24.3	& -18 18 40	&	311.86	&	 44.09	        &   EP      \\
207129  &   107649  &                       &   5.58    &   0.60    &   G2V             &           &   16.0    & 21 48 15.8    & -47 18 13 &   350.88  &   -49.11          &   ($^j$)     \\
\hline
\multicolumn{13}{l}{\textit{Active Stars}} \\
10361	&	7751 A	&	p Eri B	            &	5.80	&	0.89	&	K5Ve$^k$        & 11.22     &	 7.8	& 01 39	47.6	& -56 11 47	&	289.60	&	-59.66	        &  \\
22049	&	16537	&	$\epsilon$ Eri$^l$	    &	3.73	&	0.88	&	K2V	            &           &	 3.2	& 03 32	54.8	& -09 27 30	&	196.84	&	-48.05	        &   BY, EP  \\
26965	&	19849	&	$\omicron^2$ Eri	&	4.43	&	0.82	&	K0.5V	        &           &	 5.0	& 04 15	16.3	& -07 39 10	&	200.75	&	-38.04	        &   Fl      \\
61421	&	37279	&	$\alpha$ CMi, Procyon &	0.37	&	0.42	&	F5IV-V+DQZ	    &  4.85     &	 3.5	& 07 39	18.1	&  05 13 30	&	213.70	&	 13.03	        &   BY      \\
131156	&	72659	&	$\ksi$ Boo	        &	4.59	&	0.78	&	G7Ve+K5Ve       &  7.32     &	 6.7	& 14 51	23.4	&  19 06 02	&	 23.09	&	 61.35	        &   BY      \\
131977	&	73184	&		                &	5.72	&	1.11	&	K4V	            &           &	 5.8	& 14 57	28.0	& -21 24 56	&	338.24	&	 32.67	        &   BY      \\
154417	&	83601	&	V2213 Oph	        &	6.01	&	0.58	&	F8V	            &           &	20.7	& 17 05	16.8	&  00 42 09	&	 20.77	&	 23.78	        &   BY      \\
165341	&	88601	&	70 Oph	            &	4.03	&	0.86	&	K0V	            &           &	 5.1	& 18 05	27.3	&  02 30 00	&	 29.89	&	 11.37	        &   BY      \\
191408	&	99461	&		                &	5.32	&	0.87$^m$&	K2.5V+M3.5	    &  5.62     &	 6.0	& 20 11	11.9	& -36 06 04	&	  5.23	&	-30.92	        &           \\
192310	&	99825	&	GJ 785, 5 G Cap	    &	5.72	&	0.91	&	K2V	            &           &	 8.9	& 20 15	17.4	& -27 01 59	&	 15.63	&	-29.39	        &   Var, EP \\
\hline
\end{tabular}
\begin{flushleft}
a - The separation of the listed companion from the primary in seconds of arc. \\
b - BY: BY Dra variable, Fl: Flare star, Var: variable, PMS: Pre-Main Sequence, HD: Hot Dust, EP: Exoplanet host system, LP: Low polarization standard. All notes come from SIMBAD with the following exceptions: $\beta$ Vir \citep{bailey10}, $\tau$ Cet \citep{difolco07}, e Eri \citep{ertel14}, Procyon \citep{schaaf08}, $\epsilon$ Ind \citep{clarke10}, Exoplanets from the NASA Exoplanet Archive \citep{akeson13}. \\
c - Spectral type is for the combined light. The A and B components have V magnitudes of 5.61 and 6.49, and $B-V$ values of 0.61 and 0.81 respectively. \\
d - Separation from \citet{giuricin84}. \\
e - Listed in SIMBAD as a spectroscopic binary, but \citet{duquennoy91} suggest otherwise. \\
f - A second companion with spectral type M1.5V is separated by 40.09$\arcsec$. \\
g - A wide (416$\arcsec$) binary companion system consists of two brown dwarfs: $\epsilon$ Ind Ba (T1) and $\epsilon$ Ind Bb (T6) \citep{mccaughrean04}. \\ 
h - $\epsilon$ Ind A is a candidate for having an exoplanetary companion with a period of 30 yr \citep{zechmeister13}. \\ 
i - Companion has a V magnitude of 11.8. \\
j - HD 207129 is listed as a pre-main sequence star in SIMBAD, however its age is given elsewhere as 1.5--3.2 Gyr \citep{marshall11}. \\ 
k - Spectral type from \citet{glebocki80}. \\
l - In addition to being an active star, $\epsilon$ Eri is also a debris disk host. It is grouped with the active stars for reasons that will be developed through the paper. \\
m - SIMBAD $B-V$ is unreliable for this star, we have substituted data from \citet{martinez-arnaiz10}. \\
\end{flushleft}
\label{tab:stars}
\end{table*}

We selected the programme stars to cover the range of spectral types between F0 and K5. The K5 cut-off being a result of the imposed magnitude limit. We didn't otherwise aim to favour stars with any particular properties and our initial target list consisted of the brightest accessible dwarf of each spectral type according to the types assigned to the stars of the Hipparcos catalogue \citep{perryman97} in the VizieR database. An additional five K dwarfs were then added to achieve an even number of F, G and K types. Where scheduling or other constraints prevented a star of a particular type being observed, we selected another with a similar spectral type. Upon completion of our observations we added five stars observed as part of a debris disk investigation programme not yet reported (Marshall et al., in prep). Some of these stars were on our original list of most preferred targets. The additional stars met the fundamental parameters of the study, being similarly bright, falling in the required spectral type range, and having been observed to the same precision limits as the other programme stars.

The chromospheric emission at which a star is considered active is not universally defined and spans a range from mildly to very active, so the small number of stars surveyed here have been separated into just two groups according to their activity levels. We observed ten stars that we could find classified as `active' in the literature. These included several BY Dra variables, stars with emission line spectral types, a flare star, and the K dwarfs HD 191408 and GJ 785, the latter of which is listed in SIMBAD as `Variable'. The classifications weren't always consistent. \citet{martinez-arnaiz10} classified HD 191408 as active but noted that it had been classified as inactive by other authors. Similarly \citet{jenkins06} describes GJ 785 as active but \citet{martinez-arnaiz10} describes it as inactive. Similarly, Procyon's status as an active star is somewhat controversial, it is described as an active star by \citet{huber11} through photometric and RV analysis but this is not supported by the classification given by \citet{martinez-arnaiz10} in their spectroscopic work. For the purposes of this work we refer to all of these stars collectively as active stars. As a result of the selection criteria and the way the programme was compiled, roughly even numbers of ordinary FGK dwarfs (14), inactive debris disk host stars (8), and active stars (10) were observed. For reasons that will be developed through the paper we present the programme stars in these groupings in Table \ref{tab:stars} and subsequently. Note that one star, $\epsilon$ Eri, is both active and a debris disk host, and we have grouped it with the active stars\footnote{To avoid confusion please note that we have both $\epsilon$ Eri and e Eri in this survey, e Eri is in the debris disk group.}; this will also be elaborated upon later.

\subsection{Observation methods}

Our observations were made with the HIPPI (HIgh Precision Polarimetric Instrument) mounted on the 3.9-m Anglo Australian Telescope (AAT). The AAT is located at Siding Spring Observatory near Coonabarabran in New South Wales, Australia. HIPPI was mounted at the f/8 Cassegrain focus of the telescope where it had an aperture size of 6.7$\arcsec$.

HIPPI is a high precision polarimeter, with a reported sensitivity in fractional polarization of $\sim$~4.3 ppm on stars of low polarization and a precision of better than 0.01 per cent on highly polarized stars \citep{bailey15}. HIPPI achieves its high precision by utilising a Ferroelectric Liquid Crystal (FLC) modulator at a frequency of 500 Hz to negate the effects of astronomical seeing. For the observations reported here an SDSS $g^{\rm \prime}$ filter was positioned, via a filter wheel, between the modulator and a Wollaston prism that splits the light into two orthogonal polarization states, which are then recorded separately by two Photo Multiplier Tubes (PMTs). Second stage chopping, to reduce systematic effects, is accomplished by rotating the entire back half of the instrument after the filter wheel through 90$\degr$ in an ABBA pattern, with a frequency that was adjusted, but was in the range of once per 40--80 seconds. An observation of this type measures only one Stokes parameter of linear polarization. To obtain the orthogonal Stokes parameter the entire instrument is rotated through 45$\degr$ and the sequence repeated. The rotation is performed using the AAT's Cassegrain instrument rotator. In practice we also repeat the observations at geometrically redundant telescope position angles of 90$\degr$ and 135$\degr$ to allow removal of instrumental polarization. 

The effects of the background sky are removed through the subtraction of a 2$\arcmin$ separated sky measurement that is acquired at each telescope position angle an object is observed in. The duration of the sky measurements was 3 minutes per Stokes parameter. The observing, calibration and data reduction methods are described in full detail in \citet{bailey15}.

The $g^{\rm \prime}$ band was chosen for our measurements mainly because it is the standard astronomical band in which HIPPI is most sensitive, to be consistent with \citet{cotton16}, and because bluer wavelengths are more sensitive to Rayleigh scattering that is most likely to be detected from exoplanets. The $g^{\rm \prime}$ band is centred on 475 nm and is 150 nm in width, which results in the precise effective wavelength and modulation efficiency -- the polarimeter's raw measurement for 100 per cent polarized light -- changing with star colour. Table \ref{tab_eff} gives the effective wavelength and modulation efficiency for various spectral types based on a bandpass model as described in \citet{bailey15}. As all of our targets are within 30 pc, no interstellar extinction has been applied in the bandpass model. Our reported results apply the efficiency correction to each star measurement (a linear interpolation is used between the given types).

\begin{table}
\caption{Effective wavelength and modulation efficiency for different spectral types according to bandpass model.}
\centering
\begin{tabular}{lcc}
\hline
Spectral    &   Effective       &   Modulation      \\
Type        &   wavelength (nm) &   efficiency (\%) \\
\hline
B0         &   459.1            &   87.7            \\
A0         &   462.2            &   88.6            \\
F0         &   466.2            &   89.6            \\
G0         &   470.7            &   90.6            \\
K0         &   474.4            &   91.6            \\
M0         &   477.5            &   92.0            \\
M5         &   477.3            &   91.7            \\
\hline
\end{tabular}
\label{tab_eff}
\end{table}

The observations were obtained predominantly over the course of two observing runs in the first semester of 2016; the first from February 25th to March 1st, the second on June 26th. A handful of serendipitously useful observations made for other programmes during earlier runs, but so far unreported, are also presented here. These data come from runs in May, June and October of 2015. The details of the conditions during those runs can be found in \citet{cotton16} and \citet{marshall16}. 

The sky was cloudless for almost the entirety of the first semester 2016 run, with seeing that varied from around 1$\arcsec$ to on rare occasions more than 6$\arcsec$, typically being between 2 and 4$\arcsec$. The seeing was similar in the 2015 runs. Of the second 2016 run, June 26th constituted the only clear night. The seeing was similar to that typically encountered in the previous run, but the last few observations were very slightly cloud affected. The effects of cloud were removed by determining the maximum signal during an observation, and rejecting integrations that fell below a threshold of 25 per cent of that. We have previously used this routine with a lower threshold \citep{cotton16}, but raised it here because the targets are on average two magnitudes fainter.

A number of stars with known high polarizations ($\sim$~1-5 per cent) were observed during each run, and used to determine the position angle zero-point; these were HD 80558 and HD 147084 in February-March and, HD 154445 and HD 147084 in June 2016. The precision of each determination is less than 1 degree, based on the consistency of the calibration provided by the different reference stars which themselves have uncertainties of this order. A difference of $\sim$~4$\degr$ found between the two runs is related to the screw-fastening of the modulator being reset, and has been accounted for through the standard rotation formula.   

The AAT is an equatorially mounted telescope, as such we use observations of stars previously measured with negligible polarizations to determine the zero-point or telescope polarization (TP). The adopted TP in May 2015 was 35.5~$\pm$~1.4 $\times 10^{-6}$, for June 2015 it was 36.5~$\pm$~1.2 $\times 10^{-6}$ \citep{cotton16}, for the October 2015 run it was 55.9~$\pm$~1.1 $\times 10^{-6}$ \citep{marshall16}. For the two 2016 runs reported here the adopted TP was 20.6~$\pm$~1.8 $\times10^{-6}$.

The AAT's primary mirror was re-aluminised the day before the beginning of the February-March run, eliminating the possibility of re-using calibration measurements made during earlier runs, but ensuring a clean surface. Preliminary calculations found the TP to be consistent within error between the February-March and June runs, and we have previously found good agreement between runs in the same semester. Consequently we combined the calibration measurements and applied them to both 2016 runs. This means that all but seven of the measurements reported here utilise the same zero-point. We used three calibration stars for the two runs, Sirius A which is only 2.6 pc distant, and $\beta$ Hyi and $\beta$ Vir, which are at similar distances ($\sim$~10 pc) to the equatorial south and north respectively. The error weighted mean polarization was determined for each star, and then the average of the three stars adopted as the TP. The details of the individual observations are given in Table \ref{tab_tp}.

\begin{table}
\caption{Low polarization star measurements to determine telescope polarization (TP) for the February-March and June 2016 runs in the $g^{\rm \prime}$ filter.}
\centering
\begin{tabular}{llcll}
\hline
Star &  Date &     $p$ (ppm) &   \hspace{2 mm}$\theta$ ($\degr$) \\
\hline
Sirius A    &           &   18.0   ~$\pm$~  0.6 &   84.7   ~$\pm$~  1.7 \\
            &   26 Feb  &   18.3   ~$\pm$~  0.8 &	87.0   ~$\pm$~	2.4 \\
            &   26 Feb  &   18.4   ~$\pm$~  2.0 &	82.6   ~$\pm$~  6.6 \\
            &   27 Feb  &   27.7   ~$\pm$~  3.7 &   83.1   ~$\pm$~  7.2 \\
            &   28 Feb  &   17.1   ~$\pm$~  0.9 &   82.0   ~$\pm$~  2.9 \\
$\beta$ Vir &           &   20.0   ~$\pm$~  4.1 &   79.3   ~$\pm$~ 11.7 \\
            &   26 Feb  &   18.6   ~$\pm$~  6.2 &   80.9   ~$\pm$~ 19.4 \\
            &   25 Jun  &   15.9   ~$\pm$~  8.9 &   88.0   ~$\pm$~ 31.4 \\
            &   25 Jun  &   24.9   ~$\pm$~  6.7 &   74.4   ~$\pm$~ 15.6 \\
$\beta$ Hyi &           &   24.1   ~$\pm$~  3.7 &   85.7   ~$\pm$~  9.0 \\
            &   25 Jun  &   19.0   ~$\pm$~  9.0 &  106.8   ~$\pm$~ 28.6 \\
            &   25 Jun  &   26.6   ~$\pm$~  4.1 &   83.5   ~$\pm$~  8.8 \\
\hline
Adopted TP  &           &   20.6   ~$\pm$~  1.8 &   83.3   ~$\pm$~  5.2 \\
\hline
\end{tabular}
\label{tab_tp}
\end{table}

\section{Results}
\label{sec:results}

\begin{table*}
\caption{HIPPI linear polarization measurements.}
\centering
\begin{tabular}{llcccrrrrrr}
\hline
Name            & HD        & Obs. & \multicolumn{1}{c}{Date} & \multicolumn{1}{c}{UT} & \multicolumn{1}{c}{Exp.} & \multicolumn{1}{c}{$q$} & \multicolumn{1}{c}{$u$} & \multicolumn{1}{c}{$p$} & \multicolumn{1}{c}{$\theta$}  & \multicolumn{1}{c}{$\hat{p}^a$} \\
                &           &       &  (dd/mm/yy)   &  (hh:mm)   &  (s)     & \multicolumn{1}{c}{(ppm)} & \multicolumn{1}{c}{(ppm)} & \multicolumn{1}{c}{(ppm)} & \multicolumn{1}{c}{($\degr$)} & \multicolumn{1}{c}{(ppm)}     \\
\hline
\multicolumn{11}{l}{\textit{Ordinary FGK Dwarfs}} \\
p Eri A         &   10360	&	1	&	01/03/16    &   10:03    &   1480	&	 -6.6~$\pm$~9.8	    &	  1.1~$\pm$~10.5	            &	 6.7~$\pm$~10.1	                &	 85.4~$\pm$~38.5	    &    0.0  \\
$\tau^6$ Eri    &   23754	&	1	&	26/02/16    &   11:56    &    640	&	 -1.3~$\pm$~6.9	    &	-16.8~$\pm$~\phantom{0}6.7	    &	16.8~$\pm$~\phantom{0}6.8	    &	132.8~$\pm$~13.2	    &   15.4  \\
$\pi^3$ Ori     &   30652	&	1	&	28/02/16    &   10:10    &    640	&	 -3.5~$\pm$~4.6	    &	 -6.2~$\pm$~\phantom{0}4.6	    &	 7.1~$\pm$~\phantom{0}4.6	    &	120.4~$\pm$~23.1	    &    5.4  \\
$\gamma$ Lep    &   38393	&	1	&	26/02/16    &   12:34    &    640	&	  0.3~$\pm$~5.6	    &	 -8.2~$\pm$~\phantom{0}5.5	    &	 8.2~$\pm$~\phantom{0}5.5	    &	136.0~$\pm$~24.1	    &    6.0  \\
9 Pup           &   64096	&	1	&	29/02/16    &   13:01    &   1280	&	  6.9~$\pm$~6.6	    &	 -8.0~$\pm$~\phantom{0}6.6	    &	10.6~$\pm$~\phantom{0}6.6	    &	155.5~$\pm$~22.3	    &    8.2  \\
HR 4523         &   102365	&	1	&	28/02/16    &   16:29    &   1024	&	-10.6~$\pm$~6.6	    &	  5.7~$\pm$~\phantom{0}6.6	    &	12.0~$\pm$~\phantom{0}6.6	    &	 75.9~$\pm$~19.4	    &   10.1  \\
$\beta$ Vir     &   102870$^b$&	3	&	            &            &   1920	&	  1.3~$\pm$~4.2	    &	  2.5~$\pm$~\phantom{0}4.2	    &    2.9~$\pm$~\phantom{0}4.2	    &	 31.3~$\pm$~38.2	    &    0.0  \\
                &           &       &   26/02/16    &   17:37    &    640   &     2.3~$\pm$~6.5     &     1.1~$\pm$~\phantom{0}6.5     &                           &                               &              \\
                &           &       &   25/06/16    &   08:43    &    640   &     4.1~$\pm$~9.2     &    -3.6~$\pm$~\phantom{0}8.9     &                           &                               &              \\
                &           &       &   25/06/16    &   09:10    &    640   &    -1.3~$\pm$~6.9     &     8.1~$\pm$~\phantom{0}7.1     &                           &                               &              \\
GJ 501.2        &   114613	&	1	&	25/06/16    &   10:13    &   1024	&	-21.5~$\pm$~7.7	    &	 31.0~$\pm$~\phantom{0}7.6	    &	37.7~$\pm$~\phantom{0}7.7	    &	 62.3~$\pm$~\phantom{0}5.9 &   36.9       \\
i Cen           &   119756	&	1	&	25/06/16    &   09:44    &    640	&	-16.9~$\pm$~7.4	    &	 19.8~$\pm$~\phantom{0}7.4	    &	26.0~$\pm$~\phantom{0}7.4	    &	 65.2~$\pm$~\phantom{0}8.4 &   25.0       \\
16 Lib          &   132052	&	1	&	27/02/16    &   17:22    &    640	&	  8.6~$\pm$~6.9	    &	  0.4~$\pm$~\phantom{0}6.8	    &	 8.7~$\pm$~\phantom{0}6.9	    &	  1.2~$\pm$~27.5	    &    5.3    \\
$\lambda$ Ser   &   141004	&	1	&	26/02/16    &   17:10    &    640	&	  1.4~$\pm$~8.5	    &	 12.8~$\pm$~\phantom{0}8.8	    &	12.9~$\pm$~\phantom{0}8.7	    &	 42.0~$\pm$~23.9	    &    9.5     \\
GJ 667          &   156384	&	1	&	01/03/16    &   18:01    &   2560	&	  5.2~$\pm$~7.6	    &	 -1.5~$\pm$~\phantom{0}7.7	    &	 5.4~$\pm$~\phantom{0}7.6	    &	172.2~$\pm$~37.5	    &    0.0    \\
$\psi$ Cap      &   197692	&	1	&	25/06/16    &   15:01    &    800	&	-12.9~$\pm$~7.3	    &	-13.2~$\pm$~\phantom{0}8.4	    &	18.5~$\pm$~\phantom{0}7.8	    &	112.8~$\pm$~14.1	    &   16.8    \\
$\epsilon$ Ind  &   209100	&	1	&	25/06/16    &   13:33    &   1280	&	  4.1~$\pm$~9.0	    &	 -7.7~$\pm$~\phantom{0}8.8	    &	 8.7~$\pm$~\phantom{0}8.9	    &	149.0~$\pm$~32.4	    &    0.0     \\
\hline
\multicolumn{11}{l}{\textit{Debris Disk Host Stars}} \\
$\zeta$ Tuc     &   1581	&	1	&   25/06/16    &	17:17    &    640	&	-11.0~$\pm$~6.8	    &	 11.4~$\pm$~\phantom{0}6.8	    &	15.8~$\pm$~\phantom{0}6.8	    &	 67.0~$\pm$~14.3	    &   14.3     \\
$\tau$ Cet      &   10700	&	2	&	            &            &   1920	&	  1.3~$\pm$~3.1	    &	  0.3~$\pm$~\phantom{0}3.0	    &	 1.4~$\pm$~\phantom{0}3.0	    &	  7.0~$\pm$~42.8	    &    0.0    \\
                &           &       &   26/06/15    &   18:28    &   1280   &    -0.8~$\pm$~4.1     &     8.0~$\pm$~\phantom{0}4.1     &                           &                               &                 \\
                &           &       &   20/10/15    &   14:20    &    640   &     4.2~$\pm$~4.8     &    -8.3~$\pm$~\phantom{0}4.3     &                           &                               &                 \\
e Eri           &   20794	&	1	&   29/02/16    &   12:27    &	  800	&	  2.3~$\pm$~6.5	    &	  4.6~$\pm$~\phantom{0}6.8	    &	 5.2~$\pm$~\phantom{0}6.7	    &	 31.6~$\pm$~36.2	    &    0.0     \\
$\zeta^2$ Ret   &   20807	&	1	&	28/02/16    &   11:58    &   1120	&	  8.2~$\pm$~7.9	    &	  3.9~$\pm$~\phantom{0}8.5	    &	 9.1~$\pm$~\phantom{0}8.2	    &	 12.7~$\pm$~30.1	    &    3.8     \\
$\eta$ Cru      &   105211	&	1	&	26/06/15    &   08:36    &    640	&	-16.8~$\pm$~6.2	    &	 12.1~$\pm$~\phantom{0}6.3	    &	20.7~$\pm$~\phantom{0}6.3	    &	 72.2~$\pm$~\phantom{0}9.0	    &   19.7   \\
$\eta$ Crv      &   109085	&	1	&	24/05/15    &   12:39    &    640	&	 -4.7~$\pm$~7.8	    &	  9.9~$\pm$~\phantom{0}8.0	    &	11.0~$\pm$~\phantom{0}7.9	    &	 57.7~$\pm$~25.3	    &    7.6    \\
61 Vir          &   115617	&	1	&	26/06/15    &   10:54    &    960	&	 -2.2~$\pm$~7.2	    &	 -2.4~$\pm$~\phantom{0}7.2	    &	 3.3~$\pm$~\phantom{0}7.2	    &	114.0~$\pm$~42.6	    &    0.0     \\
HD 207129       &   207129	&	1	&	26/06/15    &   19:11    &   1280	&	-27.9~$\pm$~8.1	    &	 -6.3~$\pm$~\phantom{0}8.0	    &	28.6~$\pm$~\phantom{0}8.0	    &	 96.3~$\pm$~\phantom{0}8.3	    &   27.4  \\
\hline
\multicolumn{11}{l}{\textit{Active Stars}} \\
p Eri B         &   10361	&	1	&	26/02/16    &   11:07    &   2560	&	  0.5~$\pm$~7.5	    &	-42.2~$\pm$~\phantom{0}7.4	    &	42.2~$\pm$~\phantom{0}7.5	    &	135.3~$\pm$~\phantom{0}5.1	    &   41.5  \\
$\epsilon$ Eri$^c$ &   22049 &	1	&	26/02/16    &   10:24    &    640	&	 28.4~$\pm$~5.6	    &	-12.0~$\pm$~\phantom{0}5.7	    &	30.8~$\pm$~\phantom{0}5.7	    &	168.5~$\pm$~\phantom{0}5.3	    &   30.3  \\
$\omicron^2$ Eri &  26965	&	1	&	29/02/16    &   09:51    &   1024	&	  4.5~$\pm$~6.0	    &	-19.3~$\pm$~\phantom{0}6.0	    &	19.9~$\pm$~\phantom{0}6.0	    &	141.6~$\pm$~\phantom{0}9.0	    &   18.9  \\
Procyon         &   61421	&	3	&	            &            &   1280	&	  4.7~$\pm$~1.5	    &	 -5.8~$\pm$~\phantom{0}1.5	    &	 7.5~$\pm$~\phantom{0}1.5	    &	154.5~$\pm$~\phantom{0}5.8	    &    7.3  \\
                &           &       &   20/10/15    &   18:14    &    320   &    12.7~$\pm$~3.1     &    -1.2~$\pm$~\phantom{0}3.1     &                           &                               &                      \\
                &           &       &   29/02/16    &   13:32    &    640   &     1.4~$\pm$~2.2     &   -10.8~$\pm$~\phantom{0}2.2     &                           &                               &                      \\
                &           &       &   01/03/16    &   09:25    &    320   &     3.6~$\pm$~2.6     &    -2.0~$\pm$~\phantom{0}2.7     &                           &                               &                      \\ 
$\ksi$ Boo      &   131156	&	2	&	            &            &   2304	&	 45.8~$\pm$~5.2	    &	  3.0~$\pm$~\phantom{0}5.2	    &	45.9~$\pm$~\phantom{0}5.2	    &	  1.9~$\pm$~\phantom{0}3.2	    &   45.6  \\
                &           &       &   26/02/16    &   18:07    &   1024   &    40.1~$\pm$~8.9     &    -2.6~$\pm$~\phantom{0}9.0     &                           &                               &                      \\
                &           &       &   29/02/16    &   18:07    &   1280   &    48.8~$\pm$~6.4     &     5.8~$\pm$~\phantom{0}6.4     &                           &                               &                      \\
HD 131977       &   131977	&	1	&	26/02/16    &   16:26    &   2560	&	  4.6~$\pm$~8.2	    &	 22.8~$\pm$~\phantom{0}8.0	    &	23.2~$\pm$~\phantom{0}8.1	    &	 39.3~$\pm$~10.8	    &   23.2          \\
V2213 Oph       &   154417	&	1	&	25/06/16    &   11:05    &   2560	&	  3.7~$\pm$~8.3	    &	 19.8~$\pm$~\phantom{0}8.5	    &	20.1~$\pm$~\phantom{0}8.4	    &	 39.7~$\pm$~13.8	    &   18.3          \\
70 Oph          &   165341	&	1	&	27/02/16    &   18:16    &    640	&	-29.0~$\pm$~9.4	    &	-17.3~$\pm$~\phantom{0}8.8	    &	33.8~$\pm$~\phantom{0}9.1	    &	105.4~$\pm$~\phantom{0}7.9	    &   32.5  \\
HD 191408       &   191408	&	1	&	25/06/16    &   12:30    &   1680	&	-15.1~$\pm$~9.3	    &	-20.7~$\pm$~\phantom{0}8.5	    &	25.6~$\pm$~\phantom{0}8.9	    &	117.0~$\pm$~10.7	    &   24.0          \\ GJ 785           &   192310	&	1	&	25/06/16    &   15:49    &   2560	&	-18.4~$\pm$~6.9	    &	  2.9~$\pm$~\phantom{0}6.8	    &	18.7~$\pm$~\phantom{0}6.9	    &	 85.5~$\pm$~11.6	    &   17.4          \\
\hline
\end{tabular}
\begin{flushleft}
a - $\hat{p}$ is debiased polarization, see the text of Section \ref{sec:results} for details. \\
b - $\beta$ Vir was used as a low polarization standard. \\
c - $\epsilon$ Eri also hosts a circumstellar debris disk. \\
\end{flushleft}
\label{tab:results}
\end{table*}

Table \ref{tab:results} gives the result for each star observed, as well as duplicate measurements of the same star below the aggregate parameters. The magnitude of linear polarization, $p$, is calculated for each star in column 7 in the usual way from normalised linear Stokes parameters $q$ and $u$: \begin{equation} \label{eq:lp} p=\sqrt{q^{2}+u^{2}}.\end{equation} Because polarization is always positive it is standard practise to debias it to best estimate the true value of $p$ when calculating the mean of a group of stars with unrelated polarization angles. Following \citet{serkowski62} debiasing is carried out according to: \begin{equation} \label{eq:S_debias} \hat{p}\sim\left.\begin{cases} (p^2-\sigma_p^2)^{1/2} & p>\sigma_p\\ 0 & p\leq \sigma_p \end{cases}\right.,\end{equation} where $\sigma_{p}$ is the error in polarization. Column 11 in Table \ref{tab:results} gives the debiased polarization for each star. For stars with multiple measurements, $q$ and $u$ are first calculated from error weighted means of the individual $q$ and $u$ observations, with $p$, $\hat{p}$ and polarization angle ($\theta$) calculated from the means.

Polarization angles are calculated as: \begin{equation}\theta=\frac{1}{2}\arctan\left(\frac{u}{q}\right).\end{equation} The calculation of the error in polarization angle, $\sigma_{\theta}$, depends on the signal to noise ratio, $p/\sigma_{p}$. If it is large then the probability distribution function for $\theta$ is Gaussian, and 1$\sigma$ errors (in degrees) are given by \citet{serkowski62}: \begin{equation} \label{eq_S_PA_Gau_Err} \sigma_{\theta} = 28.65~\sigma_{p}/p.\end{equation} However when $p$/$\sigma_{p}~<~$4 the distribution of $\theta$ becomes kurtose with appreciable wings. In such cases Equation \ref{eq_S_PA_Gau_Err} is no longer strictly accurate and instead we make use of the work of \citet{naghizadeh-khouei93} who give precisely $\sigma_{\theta}$ as a function of $p$/$\sigma_{p}$ in their Figure 2(a).

\section{Discussion}
\label{sec:discussion}

\subsection{Preliminary statistical analysis}
\label{sec:stats}

The most basic analysis possible for identifying intrinsic polarization in this type of polarimetric survey is a straight comparison of the mean polarization of two or more groups of star systems. We have done this here for a number of different categories of objects, looking at the mean debiased polarization, $\hat{p}$, taking account of the increased interstellar polarization with distance through a simple division to give $\hat{p}/d$. 

Any such analysis is confounded to a degree by interstellar polarization. All the stars observed are within $\sim$~25 pc of the Sun, and the majority are within 10 pc. Consequently we would expect that the interstellar contribution to the total polarization of a given sample be small, and that in ppm/pc for randomly distributed samples, the contribution should be fairly consistent. Despite this, without determining the direction of interstellar polarization it cannot be subtracted, and we are left with the vector sum of intrinsic ($p_{\star}$) and interstellar ($p_i$) components. However, for a large enough sample of intrinsically polarized stars, we can expect the mean polarization to be greater than the interstellar polarization alone. Furthermore, if $p_{\star} > 2p_i$ then the total polarization will always be greater than the interstellar polarization alone. The statistics are described in more detail with the aid of diagrams in \citet{cotton16} or \citet{clarke10}.

\begin{table}
\caption{Mean polarization for primary stellar groupings.}
\centering
\tabcolsep 5 pt
\begin{tabular}{l r r r r}
\hline
Group                   &   \multicolumn{1}{c}{N}   &   \multicolumn{1}{c}{Mean}   &    \multicolumn{1}{c}{Mean}            & \multicolumn{1}{c}{$\hat{p}$/d}       \\
                        &                           & \multicolumn{1}{c}{$d$ (pc)} &  \multicolumn{1}{c}{$\hat{p}$ (ppm)}   & \multicolumn{1}{c}{(ppm/pc)}          \\
\hline
Ordinary FGK Dwarfs     &  14   &  13.1     &    9.9~$\pm$~1.9  & 0.8~$\pm$~0.1     \\
Debris Disk Host Stars  &   8   &  11.6     &    9.1~$\pm$~2.5  & 0.8~$\pm$~0.2     \\
Active Stars            &  10   &   7.3     &   25.8~$\pm$~2.2  & 3.5~$\pm$~0.3     \\
All Inactive Stars$^a$  &  22   &  12.6     &    9.4~$\pm$~1.4  & 0.8~$\pm$~0.1     \\
\hline
\end{tabular}
\begin{flushleft}
a - Includes both debris disk hosts and ordinary FGK dwarfs. \\
\end{flushleft}
\label{tab:primary}
\end{table}

In Table \ref{tab:primary} we calculated the mean polarization from the primary stellar groupings as presented in Table \ref{tab:stars}. From Table \ref{tab:primary} it is clear that active stars are more highly polarized than inactive stars. This is an important finding, and we set it aside for detailed discussion in Section \ref{sec:active}, where we examine the active stars in detail. In the remainder of this section we look for other trends in the inactive stars only. 

Table \ref{tab:primary} does not reveal any significantly different polarization for debris disk host stars compared to ordinary FGK dwarfs. In \citet{cotton16} we found slightly higher polarizations for debris disk systems, and significant polarization has been seen in a number of debris disk systems with aperture techniques \citep{hough06,wiktorowicz08}, so this is somewhat surprising. However, the stars examined here are on average much closer, meaning that in many cases the debris disk might be wholly outside HIPPI's 6.7$\arcsec$ diameter aperture, or may only have a fraction inside it. In addition, the polarization of debris disk systems is complicated and depends upon a number of parameters including disk radius, extent, and inclination, as well as the optical properties of the dust grains in the disk \citep[e.g.][]{graham07,schuppler15}. This requires an in-depth analysis on a system-by-system basis, which we carry out in Section \ref{sec:dd}, but for the remainder of this Section we make no distinction between the debris disk stars and other inactive FGK dwarfs.

Other less likely scenarios for intrinsic polarization are examined in Table \ref{tab:other}. None of the comparisons produced differences of any significance beyond 1-sigma. 

\begin{table}
\caption{Mean polarization for other groupings of inactive stars.}
\centering
\tabcolsep 5 pt
\begin{tabular}{l r r r r}
\hline
Group                   &   \multicolumn{1}{c}{N}   &   \multicolumn{1}{c}{Mean}   &    \multicolumn{1}{c}{Mean}            & \multicolumn{1}{c}{$\hat{p}$/d}       \\
(Inactive)              &                           & \multicolumn{1}{c}{$d$ (pc)} &  \multicolumn{1}{c}{$\hat{p}$ (ppm)}   & \multicolumn{1}{c}{(ppm/pc)}          \\
\hline
Single              &  15   &  12.3     &    9.7~$\pm$~1.6  & 0.8~$\pm$~0.1     \\
Binary/Multiple$^a$ &   7   &  13.1     &    9.5~$\pm$~2.9  & 0.7~$\pm$~0.2     \\
Binary in Aperture  &   3   &  14.2     &   11.1~$\pm$~4.2  & 0.8~$\pm$~0.3     \\  
\hline
Exoplanet Hosts     &   4   &  11.1     &   11.7~$\pm$~3.5  & 1.1~$\pm$~0.3     \\
Non-Exoplanet Hosts &  18   &  12.7     &    9.1~$\pm$~1.7  & 0.7~$\pm$~0.1     \\
\hline
F-type              &  10   &  15.3     &   11.5~$\pm$~2.1  & 0.8~$\pm$~0.1     \\
G-type              &   9   &  11.6     &   10.7~$\pm$~2.4  & 0.9~$\pm$~0.2     \\
K-type              &   3   &   7.0     &    0.0~$\pm$~5.2  & 0.0~$\pm$~0.8     \\
\hline
\end{tabular}
\begin{flushleft}
a - This line gives the binaries/multiples as identified in Table \ref{tab:stars}, the following line includes only those binaries contained wholly within the aperture: 9 Pup, i Cen and GJ 667. \\
\end{flushleft}
\label{tab:other}
\end{table}

If there is any material entrained between a binary pair we might expect to see a polarization signal, as is the case for young close binaries \citep{mclean80}. $\eta$ Cru is a binary debris disk system. The binary debris disk system $\epsilon$ Sgr is thought to display elevated levels polarization as a result of the secondary illuminating part of the disk, creating an asymmetry in aperture measurements \citep{cotton16c}. When we consider all the binary stars as a group, Table \ref{tab:other} does not reveal any systematic increase in polarization through such mechanisms in the FGK dwarfs we observed.

For completeness we have also examined the difference between known exoplanet hosts and non-exoplanet hosts. Particularly close hot-Jupiters have the potential to induce a detectable polarization signal \citep{seager00}. It has also been proposed that the presence of a close in giant planet induces magnetic activity in the host star -- which might induce polarization -- though an attempt to observe this effect did not produce a positive result \citep{cuntz00}. None of the systems observed are known to host a sufficiently large and close planet to enact either of these mechanisms. It is extremely unlikely that any such planet would be undiscovered in a system less than $\sim$~25 pc from the Sun (but not impossible if it were in a face-on orbit or if the system specifics make it a challenging radial velocity target). Table \ref{tab:other} indicates a slightly elevated polarization for the exoplanet host stars, but only at barest 1-sigma significance. The most plausible explanation for this level of difference in the polarization signal of the two groups is the combination of a small sample size and variability in interstellar polarization.

\citet{tinbergen81} suspected the presence of variable intrinsic polarization at the 100 ppm level in stars with spectral type F0 and later. More recently \citet{cotton16} combined their measurements with those of \citet{bailey10} to reveal greater polarizations in M-type stars at about that level. The data also suggested slightly elevated levels in F, G and K types over A-type stars. However, the later studies contained a combined total of only three dwarf stars later than A9, and the conclusions regarding later types were restricted to the giant class. In Table \ref{tab:other} we compared the polarizations of F-, G- and K-types. The table contains only inactive stars. Most of the active stars are K-type stars (with only a couple of earlier types) and if included would show much higher polarizations for K-types. As it is, all three of the inactive K-type stars have a debiased polarization of zero, which doesn't make for good statistics. The table doesn't reveal any trends with spectral type. Nonetheless we take a closer look at ordinary FGK dwarfs in Section \ref{sec:FGK}.

\subsection{Interstellar polarization}
\label{sec:interstellar}

Interstellar polarization is of interest for what it can tell us about the composition and history of the ISM and the ISMF \citep{frisch14,heiles96}. In combination with gas density studies such as those of \citet{lallement03,redfield08}, polarimetry is the best tool we have for understanding the composition of the ISM close to the Sun. The dust density of the ISM may also play a role in planet formation, and \citet{helled14} have called for the development of giant planet formation models that incorporate the initial size distribution of interstellar dust grains. Accurate dust maps will be required to test such models. Recently it has been hypothesised that the atmosphere of Mars was stripped through interactions with interstellar clouds \citep{atri16}. So, mapping interstellar polarization may also tell us about the likely habitability of planets in nearby space.

\begin{figure*}
\centering
\includegraphics[width=\textwidth,trim={0 7cm 0 0cm}]{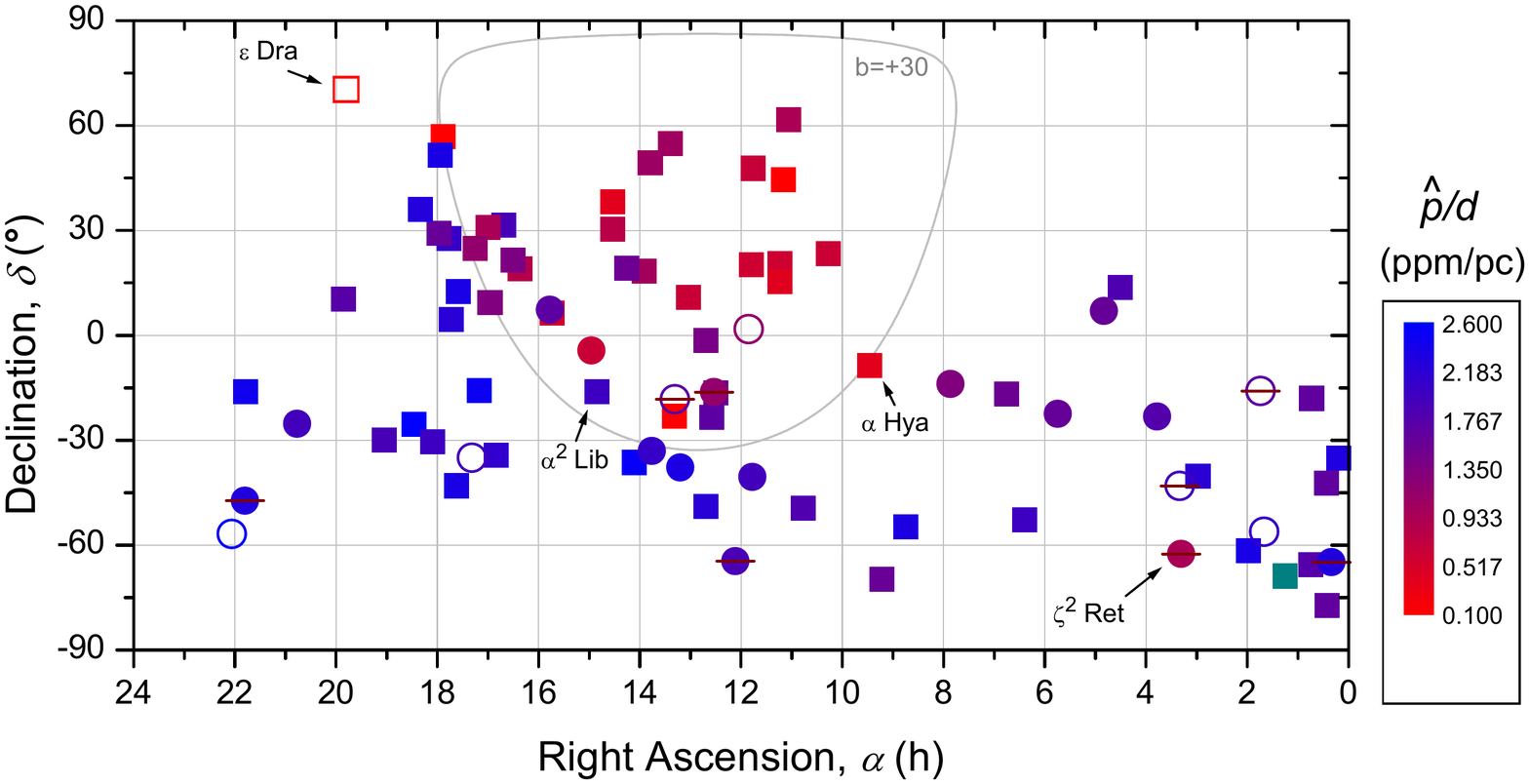}
\caption{Plot of polarization/distance ($\hat{p}/d$) vs. sky position for stars within 100 pc (most are within 50 pc). The new measurements added by this work are shown as circles. Literature measurements, shown as squares, are taken from \citep{bailey15,cotton16,cotton16b,marshall16,bailey10}. Only those stars believed to have negligible intrinsic polarization have been included. The PlanetPol values have been scaled to $g^{\rm \prime}$ according to the mean colour of the ISM determined from $g^{\rm \prime}$ and $r^{\rm \prime}$ measurements using Serkowski's Law; see \citet{marshall16} for details. Debris disk stars are indicated by a horizontal brown bar. The data point colour scale running from red to blue corresponds to 0.1 to 2.6 ppm/pc in a logarithmic fashion. Data points that debias to zero are shown as open symbols, with their colour representing the 1-sigma error. The cyan data point is HD 7693 which has a $\hat{p}/d$ value of 7.5 ppm/pc. The grey line corresponds to $b=+30\degr$. \label{fig:sky_plot}}
\end{figure*}

In Section \ref{sec:stats}, $\hat{p}/d$ amongst inactive stars was very similar no matter the exact grouping. The basic statistics therefore suggest that the inactive stars in our data set have a polarimetric signal dominated by interstellar polarization. Their measurement thus represents valuable data on the ISM close to the Sun. However, the analysis so far has only looked at groups of stars, which can lead to individual stars with significant levels of intrinsic polarization being missed. Our first step in exploring the data in this context is to repeat the exercise conducted in \citet{marshall16}. In Figure \ref{fig:sky_plot} we have plotted $\hat{p}/d$ for each inactive star along with those from the literature thought to be polarized only by the ISM with comparable errors. 

The literature data plotted represent all non-peculiar, non-debris disk, inactive A-K type stars (except $\alpha$ Tuc and $\delta$ Sgr which are believed to be intrinsically polarized) from the HIPPI \citep{bailey15,cotton16,cotton16b} and PlanetPol \citep{bailey10} bright star surveys, along with the control stars from \citet{marshall16}'s work on hot dust. We refer to these stars collectively as the \textit{Interstellar List}. A full list of the additional stars representative of the ISM and their adopted polarizations is supplied in Appendix \ref{apx:interstellar}. None of these stars belong to types known to be intrinsically polarized in the waveband of their measurement, and statistical tests very similar to those carried out in Section \ref{sec:stats} have been used to deduce only interstellar polarization \citep{bailey10,cotton16}. Where we have measurements in multiple bandpasses, the $g^{\rm \prime}$ measurement is used; for the PlanetPol observed stars we have multiplied the polarization by 1.2 in accordance with the polarimetric colour of the local ISM determined in \citet{marshall16}. It should be noted that the polarimetric colour determination, though the best available, has a very large error associated with it, and more multiband measurements of nearby stars are badly needed. A couple of the stars from the HIPPI survey have been re-observed as part of calibration procedures for later runs, and for these we have updated measurements.

The new data helps to fill out the plot compared to the \citet{marshall16} work, even whilst we exclude a number of debris disk objects included previously. Of the 22 stars newly plotted on the diagram, only one really stands out as being against trend: the debris disk system $\zeta^2$ Ret is underpolarized compared to the surrounding stars. Debatably there are other debris disk systems (marked on the plot with horizontal bars) that might also be identified as over- or under-polarized, but the apparent clumpiness of the ISM on this scale doesn't lend itself to firm identifications. We discuss the debris disk systems in more detail in Section \ref{sec:dd} after subtracting interstellar components in Section \ref{sec:isub}. However, for the remainder of this section dealing with interstellar polarization we remove all but two: e Eri -- which has a tiny infrared excess (see Section \ref{sec:dd}), and $\eta$ Crv -- where the aperture is wholly inside the cold component of the disk\footnote{There is a warm inner disk component as well, but this is dominated by small grains and likely to be very weakly polarizing at the wavelengths of interest here.}. For these reasons e Eri and $\eta$ Crv are essentially ordinary FGK dwarfs as far as HIPPI observations are concerned. Thus we have a total of 16 stars that have met the same criteria as the others on the Interstellar List, that we use to describe the local ISM.

The most striking feature of Figure \ref{fig:sky_plot} is the region of lower polarization in the northern hemisphere. This region roughly corresponds to the projected area north of +30$\degr$ galactic latitude. Though there are a few stars that appear to fall just on the wrong side of this boundary -- $\epsilon$ Dra, $\alpha^2$ Lib and $\alpha$ Hya -- which we have marked on the plot. This is not unexpected, the ISM is likely to be clumpy on this scale, and the +30$\degr$ galactic latitude line is an arbitrary boundary. Indeed, our results are not inconsistent with those of \citet{tinbergen82}, who identified what he called the `local patch' -- a region of dustier ISM centred on $l=0$, $b=-20$. The existence of this feature was brought into question by \citet{leroy93b}, but is supported by the work of \citet{frisch12}.

\subsubsection{Polarization with distance}
\label{sec:p_vs_d}

\begin{figure*}
\centering
\includegraphics[width=0.75\textwidth,trim={0 0cm 0 0cm}]{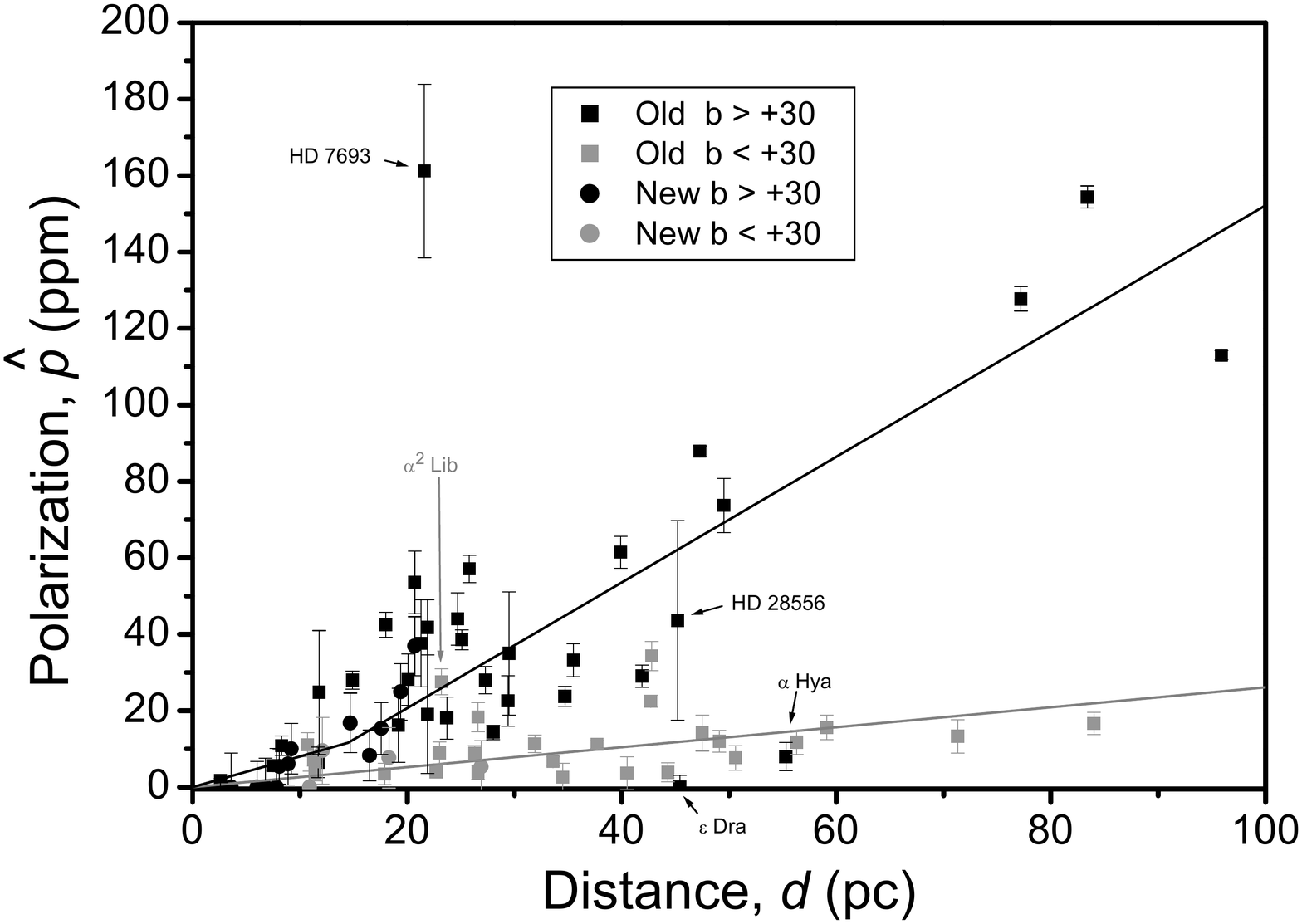}
\caption{Debiased polarization with distance for the inactive non-debris disk stars observed in this work (circles), and those from other works believed to represent interstellar polarization (squares) within 100 pc. Stars with galactic latitude greater than 30$\degr$ are plotted in grey, and the remainder in black. The lines of the same colour are linear and piece-wise linear fits to the data respectively. Stars discrepant with the apparent trends mentioned in the text are marked on the plot.}
\label{fig:p_vs_d}
\end{figure*}

\begin{figure*}
\centering
\includegraphics[width=0.75\textwidth,trim={0 0cm 0 0cm}]{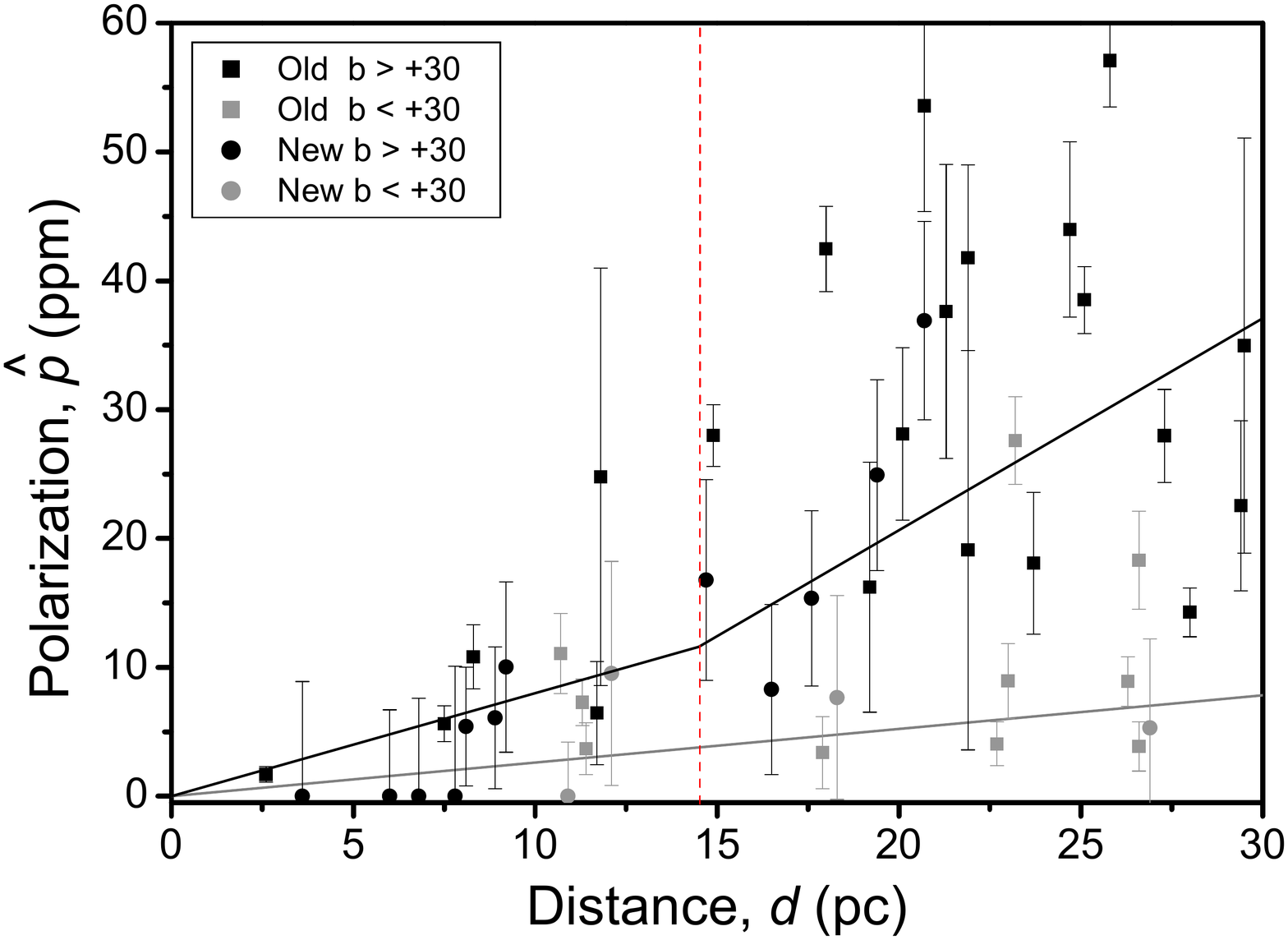}
\caption{As per Figure \ref{fig:p_vs_d} but zoomed in to within 30 pc to best show the new data stars for this work, which are all within $\sim$~25 pc. Debiased polarization with distance for the inactive non-debris disk stars observed in this work (circles), and those from other works believed to represent interstellar polarization (squares). Stars with galactic latitude greater than 30$\degr$ are plotted in grey, and the remainder in black. The lines of the same colour are linear and piece-wise linear fits to the data respectively. The red dashed line marks 14.5 pc distance.}
\label{fig:p_vs_d_zoom}
\end{figure*}

For the purpose of determining trends in polarization against distance, $\hat{p}/d$, for the groups of stars north and south of $b=+30$, we have plotted them in Figure \ref{fig:p_vs_d} in different shades -- grey for $b>+30$ and black for $b<+30$. A zoomed in version showing only stars within 30 pc is shown in Figure \ref{fig:p_vs_d_zoom}. The border region stars $\epsilon$ Dra and $\alpha$ Hya though plotted as $b<+30$ in Figure \ref{fig:p_vs_d} are used in the calculation of the $b>+30$ trend line. HD 7693 -- which appears a remarkably local phenomena -- has been excluded from the calculation, as has $\alpha^2$ Lib. We've excluded $\alpha^2$ Lib not just on account of its border status, but also because its polarization direction appears anti-aligned to surrounding stars in Figure \ref{fig:gal_plot}, leading us to suspect intrinsic polarization\footnote{Looking at this object in detail is beyond the scope of this work, but we are making follow-up observations with our mini-HIPPI instrument \citep{bailey17} designed for small telescopes.}. HD 28556 we also exclude on account of its large error. For the $b>+30$ group of stars the fitted linear trend is 0.261~$\pm$~0.017 ppm/pc. For the $b<+30$ group we initially calculate 1.318~$\pm$~0.041. These trends being fairly similar to those presented in \citet{cotton16} and \citet{bailey10}.

However, upon plotting the determined linear trend for the $b<+30$ group, it became clear that the closest stars were not well described by this simple relation. We further noted that the trend in polarization with distance for $b<+30$ stars is greater than the mean polarization with distance for inactive stars given in Table \ref{tab:primary}. Only 4 of the 22 inactive stars observed for this work belong to the $b>+30$ region, and so this does not fully explain the discrepancy. Previously \citep{cotton16} we reported that $\hat{p}/d$ seemed to be elevated between 10 and 30 pc toward the galactic south, but this elevated polarization region actually looks a bit narrower now -- closer to 15 to 25 pc. The mean distance of the inactive stars observed here is only 12.6 pc, so there are many closer stars. Examination of Figure \ref{fig:p_vs_d} suggests that within 8.5 pc of the Sun there is very little interstellar polarization. There is a very strong possibility that this is an artefact of the debiasing, given that our median precision in this study is 7.0 ppm. Models of the Loop I Superbubble (see Section \ref{sec:ismf}) place the Sun on or near its rim \citep{frisch14}. However, it does seem unlikely that the Sun would sit \textit{exactly} on the border between two regions with different $p/d$ relations, hypothesising a smoother transition between the two regions seems reasonable. According to \citep{frisch12,frisch10} the ISM has a very low density within 10 pc, and in this region is partially ionised, which indicates tight coupling of gas and dust densities, and therefore very low dust densities as well. For the $b>+30$ group of stars, if we fit a linear trend restricted to within 14.5 pc then the fit is 0.800~$\pm$~0.120 ppm/pc, which at a distance of 14.5 pc corresponds to 11.6~$\pm$~1.7 ppm; then for the $b>+30$ stars beyond that, the slope of their polarization is given by 1.644~$\pm$~0.298 ppm/pc. We adopt this relation to describe the interstellar polarization in later in Section \ref{sec:isub}. The division between the two polarization with distance regimes is marked on Figure \ref{fig:p_vs_d_zoom}.

Figures \ref{fig:p_vs_d} and \ref{fig:p_vs_d_zoom} emphasise the greater scatter amongst the $b<+30$ group compared to the $b>+30$ group. This is to be expected, given that it represents a larger volume of space. However, there may be other factors at play. Of the $b>+30$ group, a large portion are stars measured with PlanetPol at redder wavelengths and scaled to $g^{\rm \prime}$. If weak polarigenic mechanisms are stronger or more prevalent at bluer wavelengths this could explain the increased scatter in the $b<+30$ group. For instance, there are a number of K-giants amongst the literature stars plotted. Amongst them, only Arcturus (data from PlanetPol) has been identified as intrinsically polarized, and then only in the B-band \citep{kemp86,kemp87b}. However, M-giants as well as K- and M-supergiants with dust in their atmospheres show intrinsic polarization that increases as $1/\lambda$ \citep{dyck71}. This behaviour may also be present in K-giants at lower levels \citep{cotton16,cotton16b}. So it is more likely that $g^{\rm \prime}$ measurements of K giants are contaminated by small levels of intrinsic polarization. Similarly, stellar activity models show a stronger signature at bluer wavelengths \citep{saar93}, and could potentially contribute to greater scatter in the HIPPI $g^{\rm \prime}$ measurements of \textit{nominally} inactive stars.

\subsubsection{The interstellar magnetic field close to the Sun}
\label{sec:ismf}

\begin{figure*}
\centering
\includegraphics[width=\textwidth,trim={0cm 7cm 0cm 0cm}]{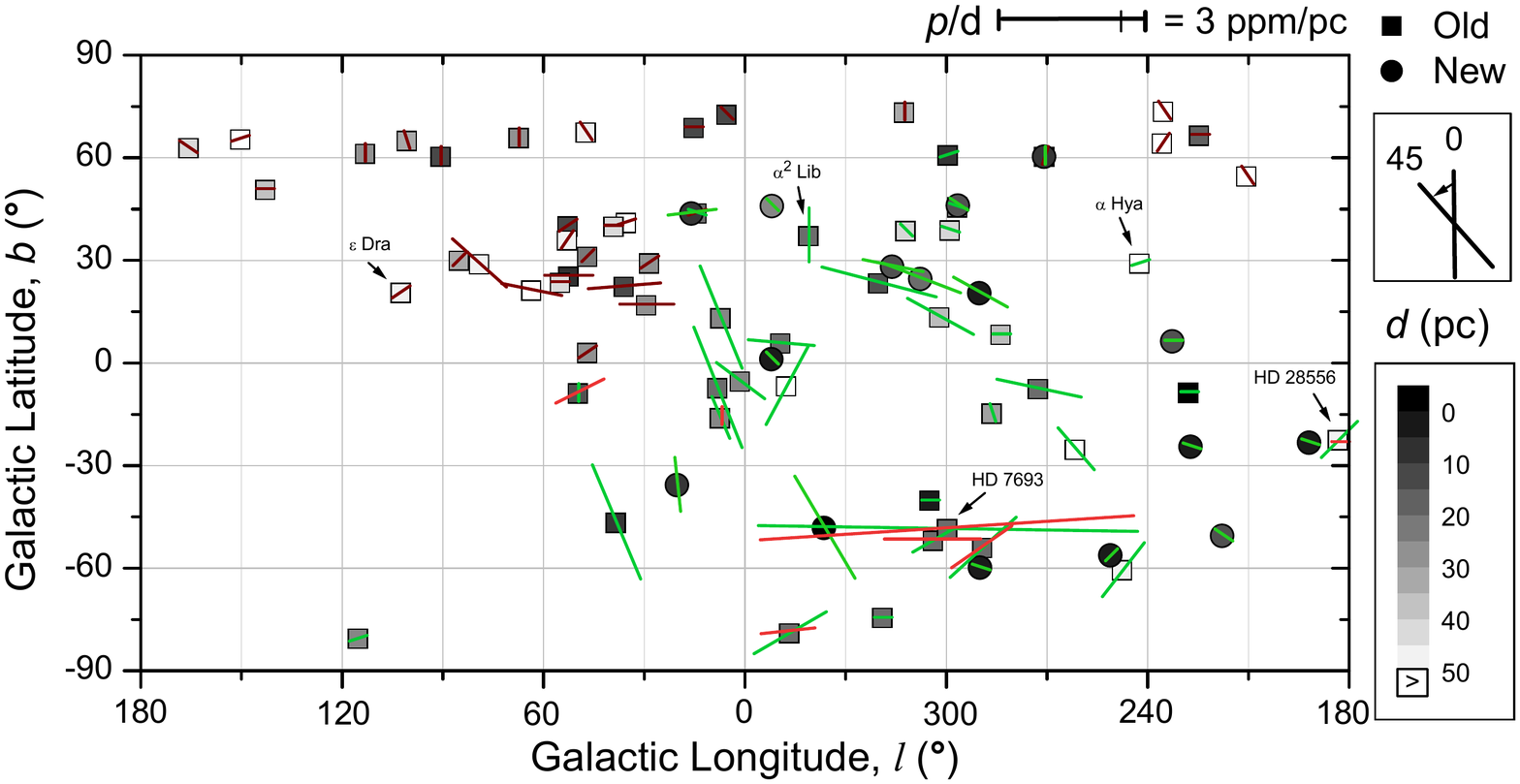}
\caption{Plot of polarization/distance ($p/d$) vs. position in galactic co-ordinates for stars within 100 pc (most are within 50 pc). The directions of the pseudo vectors give the measured galactic polarization angle, $\theta_{G}$. The new measurements added by this work are shown as circles. Literature measurements, shown as squares, are taken from \citep{bailey15,cotton16,cotton16b,marshall16,bailey10}. The vector colours are representative of the bandpasses the original measurements were made in: green for $g^{\rm \prime}$, red for $r^{\rm \prime}$ and wine for PlanetPol's red bandpass. The minimum vector length corresponds to 0.5 ppm/pc, longer vectors are representative of the polarization with distance. \label{fig:gal_plot}}
\end{figure*}

In work examining the interstellar magnetic field it is common to plot polarization vectors in galactic co-ordinates, which we do in Figure \ref{fig:gal_plot}. Here the polarization angle has been rotated into galactic co-ordinates using the method outlined by \citet{stephens11}. In this projection the polarization angle probes the magnetic structure of the local ISM. 

Close to the Sun there are two main large scale components of the ISMF. There is a uniform large scale magnetic field aligned parallel to the galactic plane towards $l=82.8$, and a local magnetic structure known as Loop I (or the Loop I Superbubble) \citep{frisch14}. The Loop I Superbubble results from stellar winds and supernovae explosions in the ScoCen association in the last $\sim$~15 Myr \citep{deGeus92,frisch95,frisch96,heiles09}. During the expansion of the Loop I Supperbubble the ISMF has been swept up, creating a magnetic bubble like structure that has persisted through the late stages of its evolution \citep{tilley06}. If Loop I is a spherical feature, the Sun sits on or near its rim \citep{frisch90,heiles98}. Optical polarization and reddening data show that the eastern parts of Loop I, $l=3$~to~$60$, $b>0$, fall within 60 to 80 pc of the Sun \citep{santos11,frisch11}. 

\citet{frisch12,frisch15} have conducted perhaps the most comprehensive study of optical polarization close to the Sun, agglomerating the PlanetPol data with a number of other data sets going back to the 1970s. That work is ongoing with an update due shortly (P.~C. Frisch, priv. comm.). The data set we present here is far less comprehensive and using it to revisit their work is beyond the scope of this paper. However, our data do contain more observations within 50 pc of the Sun, especially at southern latitudes. Distance information for each star is encoded in a greyscale colour bar in Figure \ref{fig:gal_plot}, and it can be seen that all but a handful are within 50 pc. On this scale we do not see the ridge of the Loop I superbubble traced out by polarization vectors in the same location as other studies looking at greater distances (for comparison see Figure 7 of \citet{salter83} which traces this structure in the vectors of 50 to 100 pc stars). Our results appear fairly consistent with the direction of the local interstellar magnetic field within 40 pc determined by \citet{frisch12}. Their weighted best fit gives the position of the magnetic north pole to be $l=47~\pm~20\deg, b=25~\pm~20\deg$.

In Figure \ref{fig:gal_plot} the vectors have been rendered in colours representative of the bandpasses of the original measurements. Demonstrably there is presently insufficient overlapping data in different bands to gain a good understanding of any dispersion due to the ISM. In general though, the trends in vector direction appear to be similar for the measurements made with the different instruments.

\begin{figure}
\centering
\includegraphics[width=0.5\textwidth,natwidth=100,natheight=100]{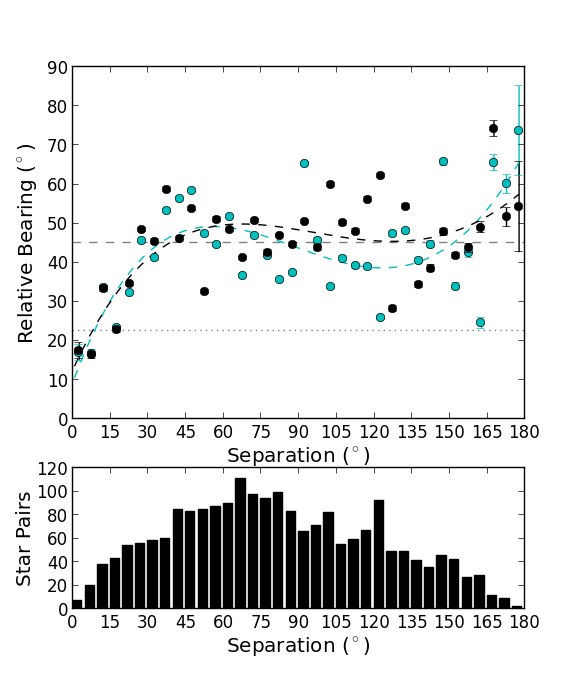}
\caption{Upper panel: Plot of the relative bearings of stars within 50 pc, binned per 5$\degr$. Data are taken from \citep{cotton16}, \citep{cotton16b}, \citep{marshall16} and \citep{bailey10}. Black data points are relative bearing calculated from polarization angles, those in cyan from galactic polarization angles. The dashed trend lines are fourth order polynomials drawn to guide the eye. Lower panel: A histogram showing the number of star pairs binned per point in the upper panel.\label{fig:bearing}}
\end{figure}

Although it is impractical to plot the polarization angle error, it is worth noting that the errors are larger for the closer stars on account of them being less polarized by the ISM. Despite this there is a high degree of coincidence in the polarization angles of stars with their 2d-neighbours, and no obvious discrepancy with distance.  It is a common practice in astronomical polarimetry to make measurements of nearby control stars to determine the interstellar polarization, and then subtract this from the target's measured polarization \citep{clarke10}. Near the Sun it can be at times very difficult to identify sufficiently close control stars. With this in mind we have endeavoured to determine the scale over which the local interstellar magnetic field rotates the angle of interstellar polarization. To do this we consider every star within 50 pc plotted in Figure \ref{fig:gal_plot}. We then measure the absolute difference in polarization angle between each star and every other star; which gives a value between 0$\degr$ and 90$\degr$ -- for simplicity we refer to this as the relative bearing. We then place the data into bins for each 5$\degr$ separation between pairs of stars, taking the error weighted mean for each bin. The result, both for galactic polarization angle and polarization angle (i.e. equatorial co-ordinates) is plotted in Figure \ref{fig:bearing}.

Statistically, for an ensemble of unrelated stars the mean relative bearing will be 45$\degr$. For neighbouring stars the interstellar magnetic field orientates them similarly, and we can see from Figure \ref{fig:bearing} that within 35$\degr$ separation there is a fairly smooth increase in relative bearing with separation. Fitted fourth order polynomials are plotted as indicative of the data trend. The trend lines don't pass through zero -- but closer to 12.5$\degr$ -- for which there are probably a number of contributing factors. Firstly there are only 8 pairs in the first (0 to 5$\degr$) bin, and 25 in the second bin; the individual measurements also have some errors associated with them. Magnetic turbulence may also be a factor. \citet{frisch12} have previously determined a trend in polarization angle rotation with distance for the PlanetPol data within 16 to 20 hours right ascension. Their best fit trend had a standard deviation about the line of 23$\degr$ attributed to magnetic turbulence. The actual trend they determined amounted to $\sim$~0.25$\degr$/pc. Which over the 50 pc range of the data plotted here amounts to 12.5$\degr$. All of these factors, together with any unidentified intrinsic or local effects will be contributing to the deviation from zero.

In Figure \ref{fig:bearing} is that there is a large amount of scatter around 45$\degr$ relative bearing at larger separations. There is also a difference between using the polarization angle and the galactic polarization angle at large separations. The measure trend line calculated using the galactic polarization angle appears negatively correlated at the largest separations. This may be attributed to the unevenness of the distribution of data along with large scale symmetry associated with the galactic magnetic field.

\subsubsection{A simple method for determining the angle of interstellar polarization}
\label{sec:interstellar_angle}

We have a determination of the magnitude of interstellar polarization with distance from Section \ref{sec:p_vs_d}. To carry out a vector subtraction of interstellar polarization for each star we also need a determination of the angle of interstellar polarization for each star. Figure \ref{fig:bearing} shows there is a fair degree of correspondence between the polarization angles of neighbouring stars that we might use to make such a determination. In this section we trial a number of different methods for determining the angle of interstellar polarization. To do this we make use of the \textit{Interstellar List} which includes all the same stars as Figure \ref{fig:bearing} within 50 pc. For each method we calculate the difference in the angle determined for each star in the Interstellar List with the angle actually measured, and decide on the best method using the mean difference (Table \ref{tab:angle_methods}). A brief description of each method follows:
\begin{enumerate}
\item \textit{Mean PA Method}: The angle for each star is the mean of the polarization angles of all other stars in the Interstellar List within 35$\degr$ separation.
\item \textit{Mean PA Separation Weighted Method}: As for the Mean PA Method, but the individual polarization angles are weighted by angular separation as: \begin{equation}Wt=(1-s_a/35),\end{equation} where $s_a$ is the angular separation in degrees. (The weighting approaches zero at 35$\degr$ separation.)
\item \textit{Mean PA Error Weighted Method}: As per the Mean PA Method, but the individual angles are weighted according to the inverse of their square error.
\item \textit{Mean PA Distance Weighted Method}: As per the Mean PA Method, but the individual angles are weighted according to: \begin{equation}Wt=\left.\begin{cases}d_c/d_t & d_c>d_t\\ d_t/d_c & d_t>d_c\end{cases}\right.,\end{equation} where $d_t$ is the distance to the target star, and $d_c$ the distance to the control star from the Sun.  
\item \textit{Mean GPA Method}: As per the Mean PA Method, but the individual polarization angles are first transformed to galactic polarization angle to take the mean, before being transformed back to polarization angle.
\item \textit{Mean GPA Separation Weighted Method}: As per the Mean PA Method Separation Weighted Method, but the individual polarization angles are first transformed to galactic polarization angle to take the mean, before being transformed back to polarization angle.
\item \textit{Mean Stokes per Distance Method}: The $q$ and $u$ vectors in ppm/pc for each star in the Interstellar List were averaged in this method. This essentially weights the angles by the strength of the polarization with distance.
\item \textit{Magnetic Field Method}: Here we determine the direction of the magnetic field at each target star's sky position based on that derived by \citet{frisch12}, and assume the direction of the magnetic field lines corresponds to the polarization direction. Doing this involved transforming the lines of longitude in a magnetic field co-ordinate system to an equatorial co-ordinate system, which involved determining the longitude of the ascending node of the magnetic co-ordinate system from the plots in \citet{frisch12} as 43.33$\degr$.
\end{enumerate}
 
\begin{table}
\caption{A comparison of different methods for determining interstellar polarization angle.}
\centering
\begin{tabular}{lcc}
\hline
Method    &   Mean Difference ($\degr$)       \\
\hline
Mean PA                             &   30.3            \\
Mean PA Separation Weighted         &   29.1            \\
Mean PA Error Weighted              &   39.1            \\
Mean PA Distance Weighted           &   31.6            \\
Mean GPA                            &   30.0            \\
Mean GPA Separation Weighted        &   30.5            \\
Mean Stokes per Distance            &   50.6            \\
Magnetic Field                      &   40.1            \\
\hline
\end{tabular}
\label{tab:angle_methods}
\end{table}

Table \ref{tab:angle_methods} indicates that the Mean PA Separation Weighted method is the best, and so we adopt it in determining the interstellar polarization. This method is only slightly better than the Mean PA method, which is completely unweighted. The reason the improvement is only slight has to do with the number of stars in the Interstellar List, and on many occasions few being very close in terms of separation on the sky. This can lead to a determination being heavily weighted to one or two stars. Any star on the list could have an unidentified intrinsic component, be misaligned through magnetic turbulence, or be poorly constrained, and so it is better to average more stars. Statistically, stars with larger polarizations are more likely to have a large unidentified intrinsic component, and the especially poor perfomance of the Mean Stokes per Distance method suggests that there may be a number of these stars. 

When we tried reducing the angular separation cut-off to less than 35$\degr$ the mean difference also increased because of a reduced number of control stars per target. Similarly the Mean PA Distance Weighted method is worse because the statistical disadvantage of favouring a smaller number of control stars outweighs the distance weighting's advantage better accounting for rotation with distance. The Mean PA Error Weighted Method is much worse than the Mean PA Method. Again, this is a consequence of differing error levels effectively reducing the number of control stars over which the average is taken. 

\begin{figure}
\centering
\includegraphics[width=0.5\textwidth,trim={0.5cm 0.75cm -0.25cm 0cm}]{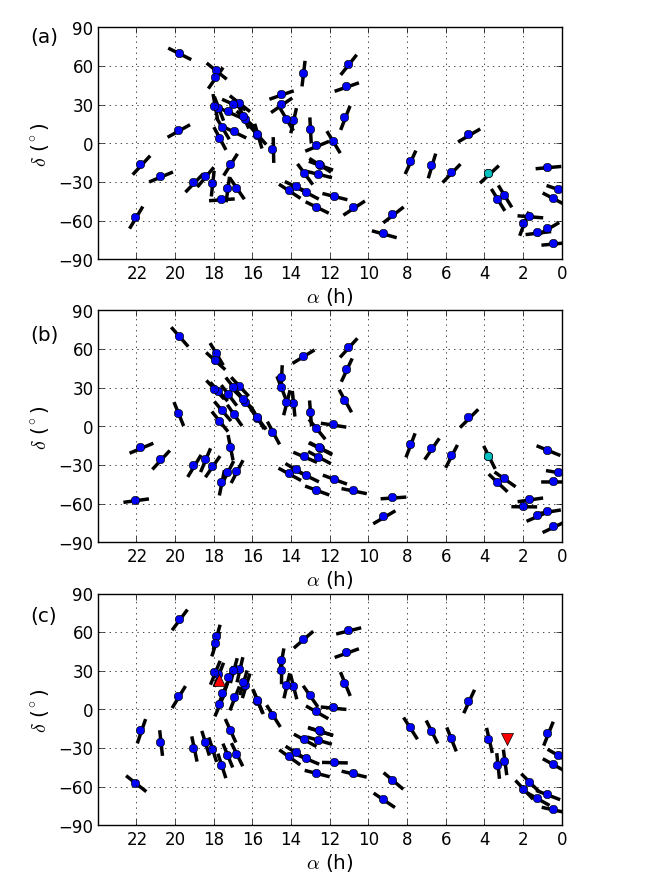}
\caption{(a) The measured polarization angles of stars representing interstellar polarization within 50 pc. (b) The polarization angles determined for the interstellar polarization from the Mean PA method. The cyan point in subplots (a) and (b) is $\tau^6$ Eri (03h~47$\arcmin$, -23$\degr$~15$\arcmin$). (c) Polarization angles determined from the Magnetic Field method, based on \citet{frisch12}. Here the magnetic north pole (up-pointing red triangle) is positioned at (265.5$\degr$, 23.0$\degr$) and the longitude of the ascending node is 43.33$\degr$; the south magnetic pole is shown as a down-pointing red triangle. \label{fig:interstellar_PA}}
\end{figure}

The galactic polarization angle methods do not do significantly better than the polarization angle methods. Unlike on larger scales the magnetic field probed by stars in nearby space probably doesn't correlate as closely to galactic co-ordinates. The Magnetic Field method might therefore be expected to do better, but doesn't on the whole. Examination of Figure \ref{fig:interstellar_PA} shows that there are actually regions of the sky where this method is doing very well, and others where it is not. One likely explanation for this is that the error in the determination of the pole position is large -- 20$\degr$ in each direction. \citet{frisch12}'s determination had the high precision PlanetPol data to work with, but little high precision data at southern latitudes. This is evident in the figure where the agreement is much better near the north magnetic pole. There are potentially other significant contributors to the interstellar polarization direction too, not just the magnetic field, including, for instance the IBEX ribbon \cite{frisch10}. In this instance however we are trying to obtain a simple method, and considering all the magnetic structure within the local ISM is beyond the scope of the present work.

\subsection{Interstellar subtraction}
\label{sec:isub}

\begin{table*}
\caption{Interstellar and intrinsic polarization components of FGK dwarfs$^a$}
\tabcolsep 5 pt
\centering
\begin{tabular}{lrccrrrrrrl}
\hline
Name            & \multicolumn{1}{c}{HD}  & $p$ &   $\theta$  &   $p_i$   &   $\theta_i$ ($\degr$)    &   $p_{\star}$   &   $\theta_{\star}$  &   $\hat{p}_{\star}$ & \\
\hline    
\multicolumn{9}{l}{\textit{Ordinary FGK Dwarfs}}   \\
p Eri A         &   13060   &   6.7~$\pm$~10.1            &   85.4~$\pm$~38.5           &    6.3    &    98.3   &    2.9    &   15.4    &    0.0    & \\
$\tau^6$ Eri    &   23754   &  16.8~$\pm$~\phantom{0}6.8  &  132.8~$\pm$~13.2           &   16.7    &   134.1$^b$ &  0.8    &   91.5    &    0.0    & \\
$\pi^3$ Ori     &   30652   &   7.1~$\pm$~\phantom{0}4.6  &  120.4~$\pm$~23.1           &    6.5    &   135.1   &    3.5    &   87.8    &    0.0    & \\
$\gamma$ Lep    &   38393   &   8.2~$\pm$~\phantom{0}5.5  &  136.0~$\pm$~24.1           &    7.1    &   152.6   &    4.5    &  105.6    &    0.0    & \\
9 Pup           &   64096   &  10.6~$\pm$~\phantom{0}6.6  &  155.5~$\pm$~22.3           &   14.9    &   159.2   &    4.6    &   77.7    &    0.0    & \\
HR 4523         &   102365  &  12.0~$\pm$~\phantom{0}6.6  &   75.9~$\pm$~19.4           &    7.4    &    69.6   &    5.1    &   85.1    &    0.0    & \\
$\beta$ Vir     &   102870  &   2.9~$\pm$~\phantom{0}4.2  &   31.3~$\pm$~38.2           &    2.8    &    81.3   &    4.4    &   11.4    &    1.2    & \\
GJ 501.2        &   114613  &  37.7~$\pm$~\phantom{0}7.7  &   62.3~$\pm$~\phantom{0}5.9 &   21.7    &    63.8   &   16.0    &   60.3    &   14.1    & \\
i Cen           &   119756  &  26.0~$\pm$~\phantom{0}7.4  &   65.2~$\pm$~\phantom{0}8.4 &   19.7    &    59.8   &    7.7    &   79.7    &    2.1    & \\
16 Lib          &   132052  &   8.7~$\pm$~\phantom{0}6.9  &    1.2~$\pm$~27.5           &    7.0    &    28.2   &    7.3    &  155.6    &    2.3    & \\
$\lambda$ Ser   &   141004  &  12.9~$\pm$~\phantom{0}8.7  &   42.0~$\pm$~23.9           &    3.2    &    30.5   &   10.0    &   45.5    &    5.1    & \\
GJ 667          &   156384  &   5.4~$\pm$~\phantom{0}7.6  &  172.2~$\pm$~37.5           &    5.5    &   152.8   &    3.6    &   27.7    &    0.0    & \\
$\psi$ Cap      &   197692  &  18.5~$\pm$~\phantom{0}7.8  &  112.8~$\pm$~14.1           &   11.9    &   137.2   &   13.9    &   92.8    &   11.5    & \\
$\epsilon$ Ind  &   209100  &   8.7~$\pm$~\phantom{0}8.9  &  149.0~$\pm$~32.4           &    2.9    &    97.7   &    9.7    &  157.4    &    4.0    & \\  
&&&&&&                                                                   &   Mean $\hat{p_{\star}}$:    &    2.9 &~$\pm$~1.9  \\
\hline
\multicolumn{9}{l}{\textit{Inactive Debris Disk Systems}}   \\
$\zeta$ Tuc     &   1581    &  15.8~$\pm$~\phantom{0}6.8  &   67.0~$\pm$~14.3           &    6.9    &   107.3   &   16.2    &   54.6    &   14.7    & \\
$\tau$ Cet      &   10700   &   1.4~$\pm$~\phantom{0}3.0  &    7.0~$\pm$~42.8           &    2.9    &    84.8   &    4.2    &  178.7    &    2.9    & \\
e Eri           &   20794   &   5.2~$\pm$~\phantom{0}6.7  &   31.6~$\pm$~36.2           &    4.8    &    46.8   &    2.6    &  177.9    &    0.0    & \\
$\zeta^2$ Ret   &   20807   &   9.1~$\pm$~\phantom{0}8.2  &   12.7~$\pm$~30.1           &    9.6    &    97.0   &   18.6    &    9.8    &   16.7    & \\
$\eta$ Cru      &   105211  &  20.7~$\pm$~\phantom{0}6.3  &   72.7~$\pm$~\phantom{0}9.0 &   20.2    &    82.8   &    7.6    &   34.2    &    4.2    & \\
$\eta$ Crv      &   109085  &  11.0~$\pm$~\phantom{0}7.9  &   57.7~$\pm$~25.3           &    4.8    &    64.9   &    6.5    &   52.5    &    0.0    & \\
61 Vir          &   115617  &   3.3~$\pm$~\phantom{0}7.2  &  114.0~$\pm$~42.6           &    2.2    &    60.5   &    4.5    &  128.2    &    0.0    & \\
HD 207129       &   207129  &  28.6~$\pm$~\phantom{0}8.0  &   96.3~$\pm$~ 8.3           &   14.1    &   126.6   &   24.9    &   81.6    &   23.6    & \\
&&&&&&                                                                   &   Mean $\hat{p_{\star}}$:    &    7.8 &~$\pm$~2.9  \\ 
\hline
\multicolumn{9}{l}{\textit{Active Stars}}  \\
p Eri B         &   13061   &  42.2~$\pm$~\phantom{0}7.5  &  135.3~$\pm$~\phantom{0}5.1 &    6.3    &    90.9   &   42.5    &  139.6    &   41.9    & \\
$\epsilon$ Eri$^c$  &   22049   &  30.8~$\pm$~\phantom{0}5.7  &  168.5~$\pm$~\phantom{0}5.3 &    2.6    &   130.3   &   30.3    &  170.9    &   29.8    & \\
$\omicron^2$ Eri &  26965   &  19.9~$\pm$~\phantom{0}6.0  &  141.6~$\pm$~\phantom{0}9.0 &    4.0    &   128.5   &   16.4    &  144.6    &   15.2    & \\
Procyon         &   61421   &   7.5~$\pm$~\phantom{0}1.5  &  154.5~$\pm$~\phantom{0}5.8 &    2.8    &   158.4   &    4.7    &  152.2    &    4.5    & \\
$\ksi$ Boo      &   131156  &  45.9~$\pm$~\phantom{0}5.2  &    1.9~$\pm$~\phantom{0}3.2 &    1.7    &    22.5   &   44.6    &    1.1    &   44.3    & \\
HD 131977       &   131977  &  23.8~$\pm$~\phantom{0}8.1  &   39.3~$\pm$~10.8           &    1.5    &    40.5   &   21.7    &   39.2    &   20.1    & \\
V2213 Oph       &   154417  &  20.1~$\pm$~\phantom{0}8.4  &   39.7~$\pm$~13.8           &   21.7    &    35.5   &    3.4    &   96.7    &    0.0    & \\
70 Oph          &   165341  &  33.8~$\pm$~\phantom{0}9.1  &  105.4~$\pm$~\phantom{0}7.9 &    4.1    &    31.4   &   37.3    &  107.1    &   36.2    & \\
HD 191408       &   191408  &  25.6~$\pm$~\phantom{0}8.9  &  117.0~$\pm$~10.7           &    4.8    &   132.4   &   21.6    &  113.7    &   19.7    & \\
GJ 785           &   192310  &  18.7~$\pm$~\phantom{0}6.9  &   85.5~$\pm$~11.6           &    7.1    &   130.5   &   20.0    &   75.1    &   18.8    & \\
&&&&&&                                                                   &   Mean $\hat{p_{\star}}$:    &    23.0 &~$\pm$~2.2  \\ 
\hline
\end{tabular}
\begin{flushleft}
a - Polarization, $p$, values are given in ppm; angle, $\theta$, in degrees ($\degr$); columns 3 and 4 are the same as in Table \ref{tab:results}, $i$ subscripts denote interstellar, whilst a star ($\star$) subscript denotes intrinsic polarization. \\
b - A manual correction was made to the polarization angle determined for $\tau^6$ Eri. See the text for details. \\
c - $\epsilon$ Eri also hosts a circumstellar debris disk. \\
\end{flushleft}
\label{tab:p_components}
\end{table*}

In this section we have determined interstellar polarizations for each of the stars in our survey using the p/d relations determined in Section \ref{sec:p_vs_d}, and the Mean PA method (Section \ref{sec:interstellar_angle}). In the first instance we treat our interstellar polarization determination as a model, neglecting the model uncertainties. This allows us to take the measurement errors for $p$ as the errors in $p_{\star}$, and lets us calculate a debiased intrinsic polarization, $\hat{p}_{\star}$, for each object using Equation \ref{eq:S_debias}. We then consider the influence of uncertainties in the model parameters on a case-by-case basis -- in general this is only necessary for the furthest stars in the $b<+30$ group. 

The Mean PA method failed in an obvious way for one star out of the 32 in the survey. The polarization angle of $\tau^6$ Eri can be seen in Figures \ref{fig:interstellar_PA}(a) ($\tau^6$ Eri is marked as a cyan point) along with the nearby control stars. It appears that $\tau^6$ Eri lies to just one side of an inflection in the interstellar magnetic field, where the field lines run in near-perpendicular directions; its polarization angle matching the stars at higher declinations very well. Our model used the four nearest stars to determine a polarization angle of 26.5$\degr$, where more weight was given to the two stars on the other side of the inflection at lower declinations, the result can be seen in Figure \ref{fig:interstellar_PA}(b). To compensate we've excluded the two control stars at lower declinations from the determination, and used only the other two to produce a polarization angle of 134.1$\degr$, which is very close to matching $\tau^6$ Eri's measured polarization angle of 132.8$\degr$.

\begin{figure*}
\centering
\includegraphics[width=\textwidth,trim={0cm 2cm 0cm 0cm}]{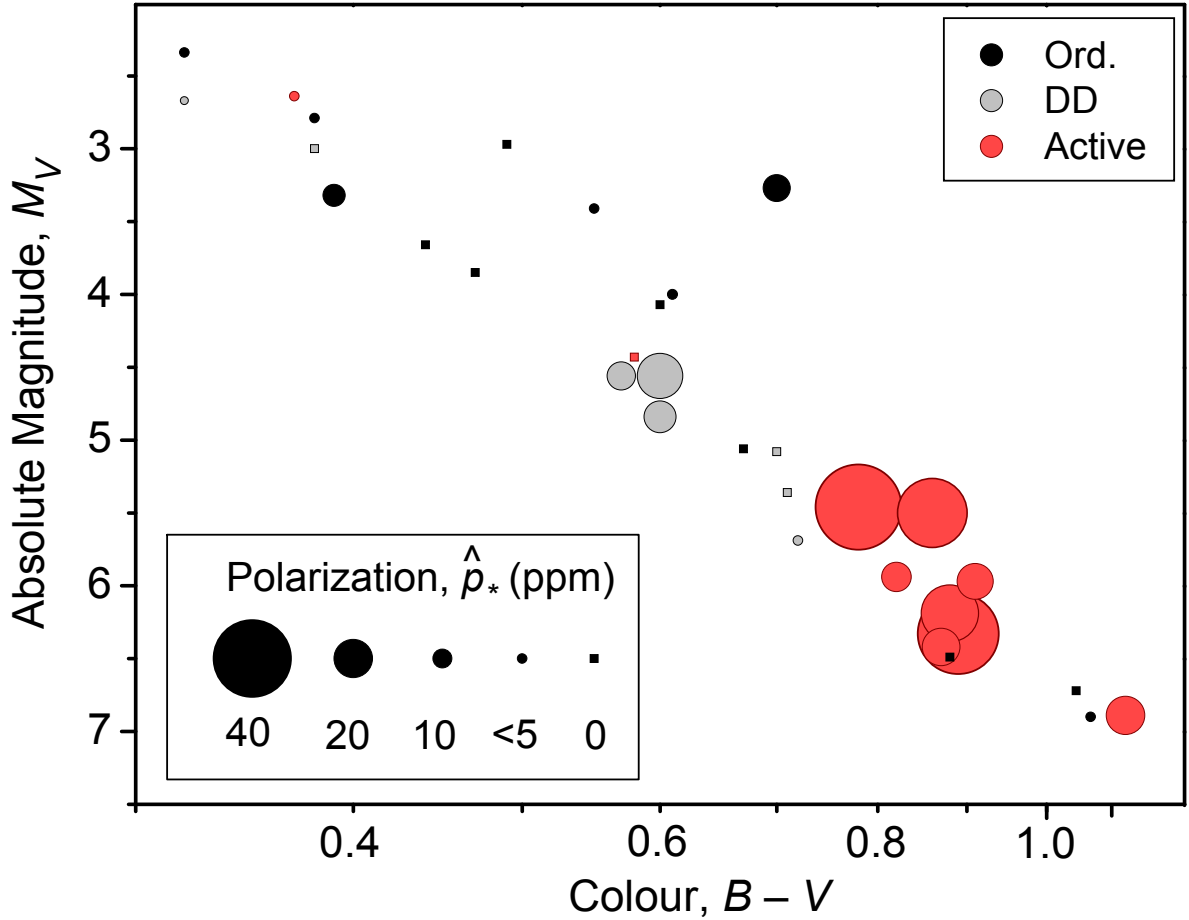}
\caption{H-R diagram showing the debiased intrinsic polarization for the FGK dwarfs of our survey. It is clear that active stars are more highly polarized, particularly those of $B-V$ colour greater than 0.75. A few debris disk systems can also be seen to have elevated polarization magnitudes. For most ordinary FGK dwarfs we calculate no significant intrinsic polarization. \label{fig:FGK_HR}}
\end{figure*}

The results of the interstellar subtraction are given in Table \ref{tab:p_components}. The magnitude of polarization of the active stars is shown to be 10 times greater than Inactive Non-Debris Disk stars. And, in contrast to the pre-interstellar subtraction result given in Table \ref{tab:primary}, the Debris Disk stars have a magnitude of polarization 1-sigma higher than the Inactive Non-debris Disk stars. Not shown are break-downs for binaries or exoplanet hosts, neither of which are significantly different to single stars or non-exoplanet hosts respectively after interstellar subtraction. In Figure \ref{fig:FGK_HR} we plot the calculated intrinsic polarizations on an H-R diagram. This serves to demonstrate that there is little intrinsic polarization to be found in F- and G-type main sequence stars, and emphasise the polarization seen in the later type active stars.

\subsection{Ordinary FGK dwarfs}
\label{sec:FGK}

The ordinary FGK dwarfs were included in our determinations of $\hat{p}/d$ in Section \ref{sec:p_vs_d} which were subsequently used in the interstellar subtraction in Section \ref{sec:isub}. However this should not be a significant impediment to identifying trends within this group of stars because the polarization angle associated with intrinsic polarization will be randomly distributed with respect to the polarization angle of interstellar polarization. Whilst the mean value of $\hat{p}/d$ can be expected to be elevated from its true value if intrinsically polarized stars are included, if $p_{\star} \sim< p_i$ the effect will be small and intrinsically polarized stars will still show up as a result of differences in angle.

\subsubsection{Outliers}

\begin{figure}
\centering
\includegraphics[width=0.5\textwidth,natwidth=100,natheight=100]{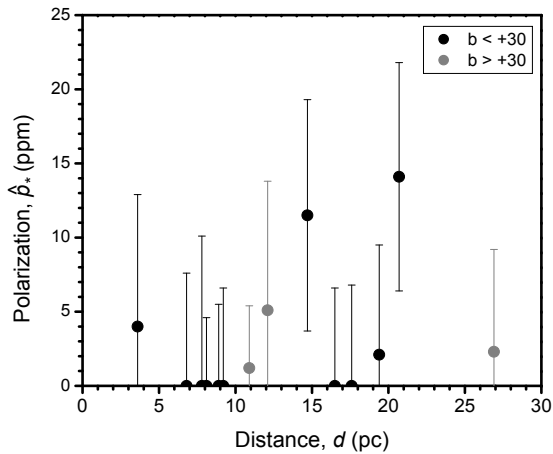}
\caption{Intrinsic polarization plotted against distance for ordinary FGK dwarfs. \label{fig:p*_vs_d}}
\end{figure}

In Figure \ref{fig:p*_vs_d}, we have plotted $\hat{p_{\star}}$ against distance for the ordinary FGK dwarfs. There is no evident trend, indicating that our interstellar subtraction is doing a reasonably good job. However in Figures \ref{fig:FGK_HR} and \ref{fig:p*_vs_d}, and in Table \ref{tab:p_components}, there are two stars that stand out with a calculated intrinsic polarization significant at around the 2-sigma level; those being the G3 dwarf GJ 501.2 and the F5 dwarf $\psi$ Cap, at distances of 20.7 and 14.7 pc, respectively. 

Seeking an explanation for the polarization of $\psi$ Cap we note that it does not have a significant infrared excess \citep{moro-martin15}, and \citet{lagrange09} has ruled out planets with a minimum mass of $m\sin{i}$ of 0.4 $M_{Jup}$ with orbital periods less than 3 days. The possibility that the polarization is a result of it being an unidentified active star is made unlikely due to its $B-V$ colour of 0.39 (refer to \ref{sec:active}). However it is interesting to note that $\psi$ Cap was the first star shown to have differential rotation using line-profile analysis and that its rotation rate is roughly 20 times that of the Sun \citep{reiners01}.

GJ 501.2 is an old (8 Gyr) and inactive star according to references within \citet{sierchio14}, so we don't expect activity to be the cause of the calculated intrinsic polarization. It may have an infrared excess at 70 $\mu$m, \citet{sierchio14} having made a detection a little below the significance they consider reliable. If correct $L_{dust}/L_{\star}$ $\sim$~10$\times$10$^{-6}$, which under ordinary circumstances is enough to account for up to (but probably less than) 5 ppm of the polarization signal. GJ 501.2 is also an exoplanet host system \citep{wittenmyer14}; where the planet has an orbital period of 10.5 yr, and a minimum mass $m\sin{i}$ of 0.48 $M_{Jup}$, which is not remotely large or close enough to expect any significant polarization signal from Rayleigh scattering \citep{seager00}. The current radial velocity limits for the system rule out planets with greater than 8 $M_{Earth}$ in orbits with semi-major axis, $a$ < 0.05 AU at 99 per cent confidence (R. Wittenmyer, priv. comm.).

The most likely explanation for the \textit{calculated} intrinsic polarization for both stars is probably interstellar polarization coupled with measurement uncertainty. Both GJ 501.2 and $\psi$ Cap are in the dustier $b<+30$ region and at a distance where the uncertainty in $\hat{p}/d$ is greater -- the 15 to 25 pc distance identified as having an elevated $\hat{p}/d$ in Figure \ref{fig:p_vs_d}. In the case of GJ 501.2 the case is particularly strong for a dustier ISM; as, from Table \ref{tab:p_components}, it can be seen that the calculated angle of interstellar polarization very closely matches the measured angle of polarization. The case is not as strong for $\psi$ Cap, but it still seems the most likely explanation.

\subsubsection{Binaries and exoplanet hosts}

Five ordinary FGK dwarfs are in multiple systems: HR 4523, 9 Pup, i Cen, GJ 667 and p Eri A. GJ 501.2, GJ 667 and HR 4523 also host exoplanets. Other than GJ 501.2, only the spectroscopic binary i Cen exhibits intrinsic polarization at any level of significance, and with a debiased polarization of only 2.1 ppm this is not worth speculating on further\footnote{$\epsilon$ Ind has a candidate planetary companion, and a debiased polarization of 4.0 ppm, but the planetary candidate is much too far from the star, and if the polarization measured is anything other than statistical noise then a very low level of stellar activity \citep{zechmeister13} is more likely to be responsible.}. These null results come despite the fact that the B components of GJ 667, i Cen and 9 Pup are inside the HIPPI aperture. Young close binaries sometimes exhibit intrinsic polarization due to gas that is entrained between the stars \citep{mclean80} or in the outer atmosphere of one of them \citep{clarke10}. Such a mechanism was invoked to try and explain the variable polarization of the young ($\sim$~70~Myr) solar type star HD 129333 \citet{elias90}. Our data suggests this phenomena is not present in any of the stars studied here.

\subsubsection{FGK stars in general}

From Table \ref{tab:p_components} there is very little, if any, intrinsic polarization in the ordinary FGK dwarfs, and no trends in $B-V$ colour or spectral type are evident. The best explanation for any \textit{calculated} intrinsic polarization here is patchiness in the dust density of the ISM in combination with statistical noise from the measurements. We therefore conclude that any increase in polarization seen in later spectral classes, such as that suspected by \citet{tinbergen81,tinbergen82} must be restricted to active stars, or higher luminosity classes as identified by \citet{cotton16}, or restricted to other wavelengths outside the $g^{\rm \prime}$ band. This means that in the $g^{\rm \prime}$ filter inactive FGK dwarfs that do not host debris disks are good probes of the local ISM, and as such make suitable interstellar calibrators for other interesting objects.

\subsection{Debris disk stars}
\label{sec:dd}

\begin{figure*}
\centering
\includegraphics[width=\textwidth,trim={0 0cm 0 0cm}]{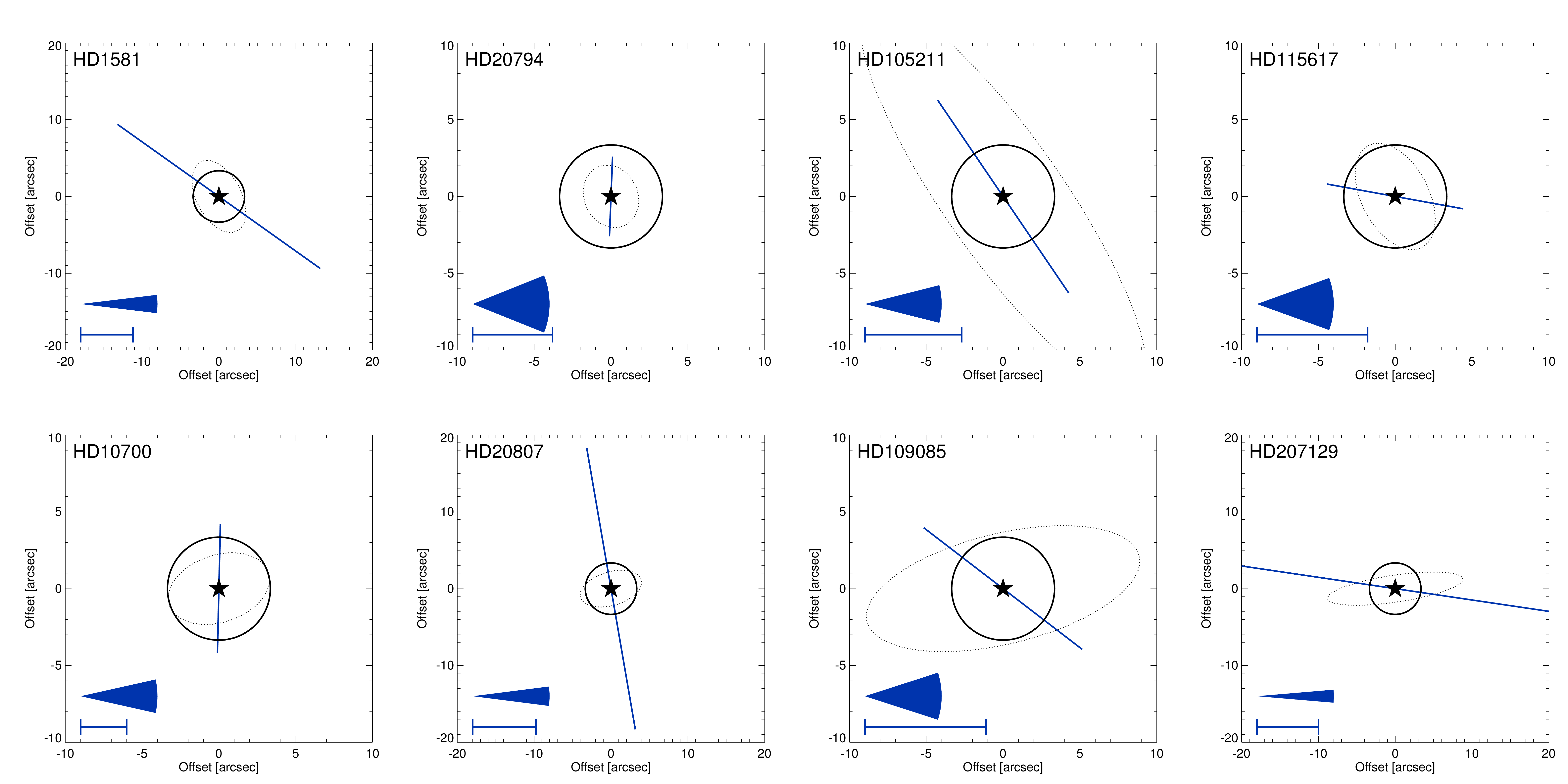}
\caption{A comparison of characteristic debris disk system parameters with the HIPPI aperture, along with polarization vectors. The dotted line shows the geometry of the disk at its characteristic radius. The solid black line shows the HIPPI aperture centred on the star. The solid blue line centred on the star shows the magnitude and angle of the polarization measured for the system (the scale marked on the individual plots also applies to the polarization vectors). The size of the 1-sigma error in polarization magnitude is shown in the bottom left hand corner of the plots by the capped blue bar, and the 1-sigma error in the angle corresponds to the angle of the blue wedges, also in the bottom left hand corner of each plot. \label{fig:dd}}
\end{figure*}

In Table \ref{tab:p_components} there are three debris disk stars with a debiased polarization of zero, one with a marginal detection -- HD 207129 -- and two with signals above the 2-sigma level. Among the factors that can influence the polarization seen from a debris disk system is its geometry with respect to the aperture. If a disk is contained wholly within the aperture then we expect the polarization vector to be aligned perpendicular to the long axis of its elliptical projection on the sky. If, on the other hand, only the edges of a system inclined edge-on are within the aperture the opposite might occur. A face-on system should be substantially unpolarized, so long as it is centred in the aperture. In order to try to make sense of this mixed bag of marginal- and non-detections we have plotted the basic system geometry of each disk system in comparison with our aperture, along with the measured polarization in Figure \ref{fig:dd}. We have also tabulated the system parameters in Table \ref{tab:dd} for reference.

\begin{table}
\caption{Debris disk properties.}
\tabcolsep 2.5 pt
\centering
\begin{tabular}{lrrrrr}
\hline
Name            & \multicolumn{1}{c}{HD} & \multicolumn{1}{c}{$r_{disk}^a$}  &   \multicolumn{1}{c}{$i_{disk}^b$}  &   \multicolumn{1}{c}{$\theta_{disk}^b$} &   \multicolumn{1}{c}{$L_{dust}/L_{\star}^a$}  \\
                &               & \multicolumn{1}{c}{($\arcsec$)}     &   \multicolumn{1}{c}{($\degr$)}     &   \multicolumn{1}{c}{($\degr$)}     &   \multicolumn{1}{c}{($10^{-6}$)} \\
\hline
$\zeta$ Tuc     &   1581	    &  3.5          &    21     &   64           &   16.0        \\
$\tau$ Cet	    &   10700	    &  3.3          &    35     &  105           &    7.8        \\
e Eri	        &   20794	    &  1.8          &    50     &    8           &    2.4        \\
$\zeta^2$ Ret	&   20807	    &  4.0          &    65     &  110           &   10.0        \\
$\eta$ Cru      &   105211      &  9.3          &    55     &   30           &   74.0        \\
$\eta$ Crv	    &   109085	    &  8.9          &    47     &  116           &   21.7        \\
61 Vir	        &   115617	    &  2.6          &    20     &   65           &   27.6        \\
HD 207129       &   207129      &  8.8          &    60     &  120           &   83.0        \\
\hline
\end{tabular}
\begin{flushleft}
a - The debris disk characteristic radius ($r_{disk}$ converted from AU) and fractional infrared excess ($L_{dust}/L_{\star}$), have been taken from the following references: $\zeta$ Tuc \citep{montesinos16,trilling08}, $\tau$ Cet \citep{lawler14}, e Eri \citep{marshall14}, $\zeta^2$ Ret \citep{eiroa13}, $\eta$ Cru (Hengst in prep.), $\eta$ Crv \citep{duchene14}, 61 Vir \citep{wyatt12}, and HD 207129 \citep{marshall11}. \\
b - The debris disk inclination ($i_{disk}$) and position angle ($\theta_{disk}$), have been taken from the following references:  $\zeta$ Tuc \citep{montesinos16}, $\tau$ Cet \citep{lawler14}, e Eri \citep{kennedy15}, $\zeta^2$ Ret \citep{eiroa10}, $\eta$ Cru (Hengst in prep.), $\eta$ Crv \citep{duchene14}, 61 Vir \citep{wyatt12}, and HD 207129 \citep{loehne12,marshall11}. \\
\end{flushleft}
\label{tab:dd}
\end{table}

Dealing with the non-detections first: e Eri (HD 20794) is the only system contained wholly within the HIPPI aperture, but it has a fractional luminosity, $L_{dust}/L_{\star}$, of 2.4$\times10^{-6}$, which in optimistic circumstances wouldn't be expected to produce a fractional polarization signal of more than 1.2 ppm. The case of 61 Vir (HD 115617) is more interesting; most of the disk is in the aperture, and it has an $L_{dust}/L_{\star}$ of 27.6$\times10^{-6}$. Modelling of the disk inferred an albedo of $< 0.31$, dominated by 1~$\mu$m grains \citep{wyatt12}. The non-detection of polarization from this system (4.5~$\pm$~7.2 ppm) is consistent with their analysis, wherein we would expect a fractional polarization at the level of $\leq$9 ppm from the whole disk. A more complex case is that of $\eta$ Crv (HD109085); it has a two component debris disk, with the inner, warm component likely delivered by bodies scattered inward from the outer disk \citep{duchene14}. The outer, cold component lies outside the HIPPI aperture, but the warm component of the disk is relatively bright, $L_{dust}/L_{\star}$ of 325$\times10^{-6}$, and lies well within the HIPPI aperture at separations down to 1 au from the star \citep{defrere15}. In this case we might infer that the dust is smoothly distributed within the HIPPI aperture, resulting in a non-detection of polarization from the system.  

We record for $\tau$ Cet (HD 10700) a very low polarization, significant only at the 1-sigma level. However, it only has an $L_{dust}/L_{\star}$ of 7.8$\times10^{-6}$ and marginally resolved Herschel observations suggest a broad, smooth disk \citep{lawler14}, so we wouldn't expect to see more polarization than is detected even with the most favourable system geometry and grain properties. Most of the $\tau$ Cet disk is contained within the aperture, and the polarization vector is roughly perpendicular to the position angle of the disk, which is what might be expected.

The $\eta$ Cru (HD 105211) system has a large infrared excess but it falls mostly outside the HIPPI aperture. The system as plotted may be misleading in this case, as the $\eta$ Cru disk shows signs of asymmetry (Hengst in prep.). However, the parts of the disk that lie within the aperture are the edges (as opposed to the ends) of the elliptical projection. The orientation of the polarization vector is consistent with what we might expect in this case.

An interesting case is $\zeta^2$ Ret (HD 20807). It is the system that initially stood out in Figure \ref{fig:sky_plot}; most of its disk lies within HIPPI's aperture, but its infrared excess is not at all large, only 10.0$\times10^{-6}$. Although our detection is formally only 2.3-sigma, a polarization of $\sim$~17 ppm is implied. The debris disk in this system is believed to be highly asymmetric \citep{eiroa13,faramaz14}. Our measurement here supports that finding. We've previously seen that asymmetry within a debris disk system can produce a larger polarization than would otherwise be expected. In our work on $\epsilon$ Sgr \citep{cotton16c} we demonstrated that a secondary component illuminating part of the disk could produce a large polarization. The polarigenic effect of a disk that has a significantly uneven dust distribution would be similar. The wide binary companion, $\zeta^1$ Ret, is separated from $\zeta^2$ Ret by 309$\arcsec$, has a similar spectral type, and no infrared excess; measurements of it would provide a very precise interstellar calibration, enabling confirmation of the polarization signal calculated here.

Another system with a 2-sigma detection and polarization greater than its infrared excess would suggest is $\zeta$ Tuc (HD 1581). In this case, the alignment of the polarization vector is not easily explainable by the system geometry. We can probably rule out an extra unsubtracted intrinsic component as the cause here because the star is quite close, only 8.6 pc. The aperture and the disk are similar sizes, so if the aperture has been positioned too far off centre we could have artificially created an asymmetry leading to a detectable polarization signal, but we don't have any reason to believe this is the case. If real, our measurement indicate some asymmetry in this disk system as well. 

HD 207129 is the only debris disk system for which we have a 3-sigma detection. It is a system that is fairly edge on ($i_{disk}=60\degr$), where the HIPPI aperture has observed the edges of the elliptical projection, but not the ends. HD 207129's infrared excess is the largest of the objects we tabulated in Table \ref{tab:dd}, so we expected a detectable polarization with a vector parallel to the position angle of the debris disk. Figure \ref{fig:dd} shows that this is close to being the case. The polarization vector is inclined $\sim$~18$\degr$ from alignment, with the 1-sigma error on our polarization angle determination being 9.7$\degr$. The polarization signal is $\sim$~30 per cent of the infrared excess, which is interesting in light of the disk's faint emission in scattered light \citep[implying a low albedo,][]{krist10} and inferred large minimum dust grain size \citep{loehne12}.

\subsubsection{Hot dust stars}

In addition to hosting a debris disks, $\tau$ Cet and e Eri are both hot dust stars \citep{difolco07,ertel14}. Hot dust is the name given to the phenomena of significant infrared excesses at near-infrared wavelengths \citep{absil13,ertel14}. The origin of the hot dust signal is still a mystery, with a leading theory being nanoscale grains \citep{su13,rieke16}. Recently \citet{marshall16} placed a strict upper limit on the polarimetric signal due to hot dust of 76 ppm in the the $g^{\rm \prime}$ band, but with a possible signal of $\sim$~17 ppm. Intriguingly \citet{ertel16} have recently published data suggesting the phenomenon may be variable. 

We measure no significant polarization for e Eri -- it is one of the least polarized objects in the survey. Either the hot dust produces no polarization in this system, or there was no hot dust present at the time of the observation.

We have two observations of $\tau$ Cet, that when averaged give the small polarization reported in Table \ref{tab:p_components}. If, on the other hand, we don't interpret the data as statistical scatter in the measurements but real variation, and do the intrinsic subtraction on each observation separately we get values of $p_{\star}=$~7.9~$\pm$~4.1 ppm, $\theta_{\star}=$~47.9$\degr$~$\pm$~18.5 on 26/06/15 and $p_{\star}=$~11.2~$\pm$~4.3 ppm, $\theta_{\star}=$~148.4$\degr$~$\pm$~13.0 four months later on 20/10/15. Particular caution needs to be taken here in interpreting these results as being due to intrinsic variability. To begin with they are hardly significant, but it needs to be said that these measurements have different TP calibrations, and at these levels small differences in the calibration will bias results in favour of variability. Nevertheless the difference is not inconsistent with the possible signal level of $\sim$~17 ppm inferred by \citet{marshall16}. We recommend long term polarimetric monitoring of stars with significant hot dust signatures.

\subsection{Active stars}
\label{sec:active}

We report here for the first time unambiguous detections of linear broadband polarization in active FGK dwarfs using an aperture technique. The stars p Eri B, $\epsilon$ Eri, $\ksi$ Boo and 70 Oph all record calculated intrinsic polarizations well in excess of 3-sigma. For $\ksi$ Boo the signal is more than 8-sigma. Additionally every active star in the survey except V2213 Oph records a greater than 2-sigma detection after interstellar subtraction. In contrast the only inactive star in the survey with a calculated intrinsic polarization greater than 3-sigma is the debris disk system HD 207129. Indeed the interstellar subtraction was hardly necessary to establish polarization in the active stars, all but one is within 10 pc and the difference between active and inactive stars was already clear in Table \ref{tab:primary}.

\subsubsection{Multiple observations of Procyon}

Procyon also records a 3-sigma signal, but we are not as confident in this detection. The reported measurement is the error weighted mean of three observations, including one in October 2015. The polarigenic mechanisms expected for active FGK dwarfs imply variable polarization. Because polarization is a pseudo-vector, averaging the Stokes parameters $q$ and $u$ from multiple measurements will underestimate the true magnitude of intrinsic polarization if it is variable. In this case an alternative formulation can be used, where each individual measurement is debiased (after interstellar subtraction), and the mean of the individual $\hat{p_{\star}}$ measurements taken. If one does this for Procyon one gets 6.3~$\pm$~1.5 ppm, which is still at the 3-sigma level, but slightly less than the result reported in Table \ref{tab:p_components}, and makes it more likely, given the small magnitude of polarization, that statistical noise or small inconsistencies in the TP calibration between 2015 and 2016 could be responsible for the detection in this case.

\subsubsection{The potential influence of $\epsilon$ Eri's derbis disk}

\begin{figure}
\centering
\includegraphics[width=0.5\textwidth,natwidth=100,natheight=100,trim={0.5cm 0cm -0.5cm 0cm}]{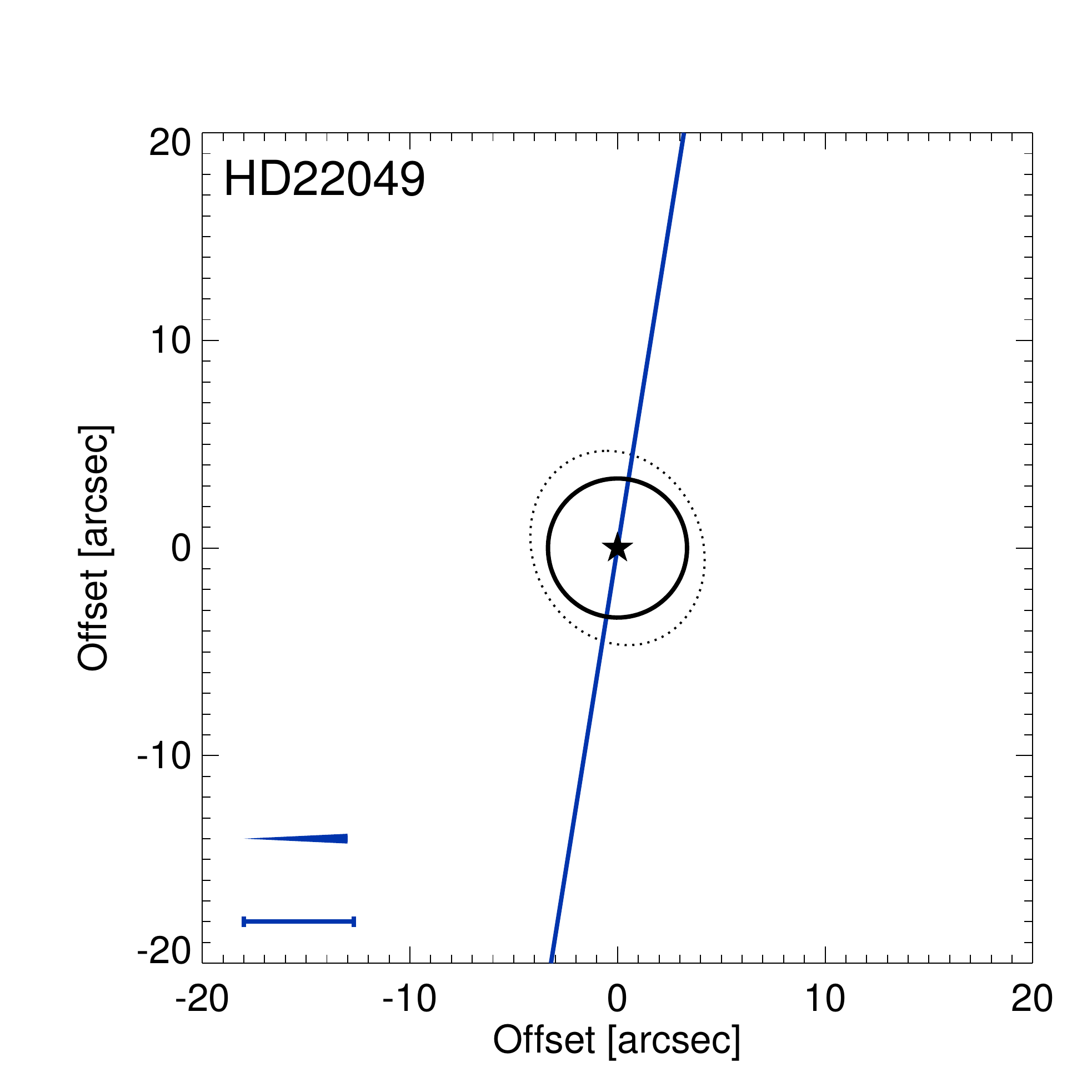}
\caption{A plot of the polarization and system geometry of $\epsilon$ Eri. The diagram is laid out as per Fig. \ref{fig:dd}, i.e. the dotted line shows the geometry of the inner belt at the outer radius of the unresolved emission shown in \citet{chavez-dagostino16}. The solid black line shows the HIPPI aperture centred on the star. The solid blue line centred on the star shows the magnitude and angle of the polarization measured for the system (the scale marked on the individual plots also applies to the polarization vectors). The size of the 1-sigma error in polarization magnitude is shown in the bottom left hand corner of the plots by the capped blue bar, and the 1-sigma error in the angle corresponds to the angle of the blue wedges, also in the bottom left hand corner of each plot. The system parameters come from \citet{marshall14,greaves14,chavez-dagostino16}: $r_{disk}=4.2\arcsec$, $i_{disk}=32\degr$, $\theta_{disk}=7\degr$. \label{fig:eps_eri}}
\end{figure}

$\epsilon$ Eri is both an active star and a debris disk system. Potentially there are components of polarization due to both of those properties. As a group the systems covered in Section \ref{sec:dd} aren't nearly as polarized as the active stars. We would therefore expect that $\epsilon$ Eri's activity is the dominant component. However, it does have a larger total infrared excess than any of the other systems: $L_{dust}/L_{\star}=107.6\times10^{-6}$. The excess is mainly due to the outer cold component, the inner warm belt around 3 au has an excess of $33\times10^{-6}$ \citep{backman09}.  We've plotted the system's parameters (for the inner belt) and polarization as we did for the other debris disk systems in Figure \ref{fig:eps_eri}. It can be seen that the characteristic radius of the belt falls just outside the HIPPI aperture. The inner system is potentially awash with dust, and no distinction can currently be made between broad or narrow architectures for the debris belts in the inner part of the system \citep{chavez-dagostino16}. The disk thus possesses an inner component that falls within the HIPPI aperture. Perhaps most importantly, the disk/belt has a near face-on inclination. As a result we would expect little contribution to the polarization due to the symmetry within the aperture. We therefore attribute the polarization seen to the star's activity.

\subsubsection{Scattering mechanisms}
\label{sec:scattering}

Recently \citet{kostogryz15b} have made calculations of limb polarization to expect in FGK dwarfs and then used this result \citep{kostogryz15} to determine the limb polarization to expect from selected (exoplanet host) FGK dwarfs due to much smaller star spots; the result being $\sim$~2-3 ppm for a spot covering 1 per cent of the stellar disc. For this mechanism to explain our results spot filling factors would have to far exceed that level. Some time prior to this \citet{saar93} tabulated the results of models estimating the maximum magnitude of limb polarization from Thompson and Rayleigh scattering that might be caused by stellar activity. Thompson scattering could not explain the magnitude of polarization we see here. 

Rayleigh scattering under optimal conditions can explain or come close to explaining our measurements. Spot sizes with filling factors of around 18 per cent optimally positioned on the surface (which represent the conditions for maximum polarization) could produce some of the levels of polarization we see. \citet{kostogryz15b}'s models indicate that linear polarization falls off quite rapidly away from the stellar limb, so \citet{saar93}'s tabulated values represent rare best-case scenarios. However, in low mass active stars the required level of spot coverage is possible \citep{jackson13}. \citet{saar93}'s specific calculations for Procyon and $\epsilon$ Eri produce 5 ppm and 33 ppm in B-band under these conditions respectively; in $g^{\rm \prime}$ their plots imply it should be about two-thirds of that, which is a little less than we measured, and a fair bit less than \citet{kochukhov11} found from spectropolarimetric measurement of $\epsilon$ Eri. The equivalent figure for $\ksi$ Boo in $g^{\rm \prime}$ band is $\sim$~100 ppm. \citet{toner88} have developed a model for $\ksi$ Boo based on spectroscopic observations that gives filling factors of 10~$\pm$~5 per cent for a feature at a latitude of 55$\degr$~$\pm$~8$\degr$. So, our measurement fits the prediction for Rayleigh scattering in this instance. 

However, the geometrical requirements for a Rayleigh scattering solution makes this mechanism seem less likely, since multiple spots sub-optimally positioned will have their effects begin to cancel out. $\ksi$ Boo has a rotation period of 6.43~$\pm$~0.01 days \citep{toner88}. We made two measurements of it exactly 3 days apart. Those measurements are not significantly different, but a change in $\theta_{\star}$ of only $\sim$~5$\degr$ is implied. If the polarization is to be attributed to a single starspot (or single patch of spots) this should not be the case unless we assume the most contrived possible combination of timing and geometry or the star was rotating pole-on -- which is not consistent with determining a rotation period from photometry.

\subsubsection{Magnetic fields}
\label{sec:mf}

A more likely polarigenic mechanism for late dwarfs is differential saturation. Active FGK dwarfs manifest net global fields of several to tens of Gauss \citep{marsden14}. Through the Zeeman effect, these fields manifest as circular polarization that is readily detected with spectropolarimetry \citep{jeffers14,morgenthaler12,fares10}. Weaker linear polarization will result from the same processes \citep{donati09}. Because it is weaker, and the line profiles more difficult to model only recently have \citet{kochukhov11} managed to detect linear polarization in an FGK dwarf using spectropolarimetry. At present there are no model predictions for broadband polarization based on the global fields of active FGK dwarfs. There are, however, older models for the broadband polarization to expect from the kG fields associated with starspots \citep{landideglinnocenti82,leroy90,huovelin91,saar93,stift97}. 

\citet{saar93}'s models produce roughly an order of magnitude greater polarization for differential saturation than Rayleigh scattering for the same spot size which, if one considers an uneven distribution of spots, matches better with what we see here. If we were to interpret our results in terms of the starspot models of \citet{huovelin91,saar93} the polarization magnitudes suggest spots with filling factors of $\sim$0.25 per cent for the less active stars to 2.0 per cent for $\ksi$ Boo. However, present mapping work using circular polarization points to significant cancellation of small-scale structure by features of opposite polarity. In light of this shift in understanding since the models were developed, their quantitative predictions are unlikely to be instructive.

Qualitatively, there are two predictions of differential saturation testable with our data. The first is that polarization increases for later stellar types (in the temperature range 4000 to 7000 K) \citep{saar93} (or with $B-V$ colour \citep{patel16}). This comes about because of increased line-blanketing in later types. The behaviour is complicated because lines are not evenly distributed, and with over-saturation net polarization is reduced. The second prediction is that for more active stars -- those with stronger magnetic fields and/or greater degrees of micro-turbulence -- have enhanced Zeeman splitting which increases polarization \citep{stift97,patel16}. These predictions are complicated by considerations of geometry \citep{huovelin91,tinbergen81} and wavelength dependence \citep{saar93,patel16}. The geometrical considerations are the most difficult to parse given a modern view of magnetic field structure in these stars, and we have neglected it here. Regarding wavelength dependence, \citet{saar93} made specific calculations for standard Johnson bands, whereas our measurements are made in the SDSS $g^{\rm \prime}$ band, which has an effective wavelength between B- and V-bands. Their trends are similar for B- and V-bands though, with V-band polarization predicted to be roughly two fifths to one quarter of that in B-band depending on the stellar parameters. 

\begin{table}
\caption{Selected properties of active stars.}
\tabcolsep 5 pt
\centering
\begin{tabular}{lrccc}
\hline
Name   & \multicolumn{1}{l}{HD} &   $B-V$   &   Abs Mag &   Activity$^a$  \\
                    &           &           &   (V)     &   $log(R'_{HK})$ \\
\hline
p Eri B	            &   10361	&	0.89	&   6.33    &   -4.94         \\
$\epsilon$ Eri	    &   22049	&	0.88	&   6.19    &   -4.62         \\
$\omicron^2$ Eri	&   26965	&	0.82	&   5.94    &   -5.38         \\
Procyon             &   61421	&	0.42	&   2.64    &   -4.75         \\
$\ksi$ Boo	        &   131156	&	0.78	&   5.46    &   -5.07         \\
HD 131977	        &   131977	&	1.11	&   6.89    &   -4.63         \\
V2213 Oph	        &   154417	&	0.58	&   4.43    &   -4.50         \\
70 Oph	            &   165341	&	0.86	&   5.50    &   -4.86         \\
HD 191408           &   191408	&	0.87$^b$&   6.42    &   -5.39         \\
GJ 785       	&   192310	&	0.91	&   5.97    &   -4.88         \\
\hline
\end{tabular}
\begin{flushleft}
a - The activity index comes from \citet{martinez-arnaiz10} for all stars listed except for GJ 785 which is the mean of two values reported by \citet{jenkins06}, and Procyon for which we have taken the S-index value form \citet{hempelmann16} and converted it to log(R$^{\prime}_{HK}$) using the relations for dwarf stars of \citet{middelkoop82} and \citet{noyes84} as related by \citet{schroeder09}. Further, it should be noted that \citet{noyes84}'s relation is strictly only valid for $B-V > 0.44$, and that Procyon falls just outside this range (0.42), and so the log(R$^{\prime}_{HK}$) value obtained is not as reliable as for the other stars listed.\\
b - SIMBAD $B-V$ is unreliable for this star, we have substituted data from \citet{martinez-arnaiz10}. \\
\end{flushleft}
\label{tab:active}
\end{table}

\begin{figure*}
\centering
\includegraphics[width=0.75\textwidth,trim={0 0cm 0 0cm}]{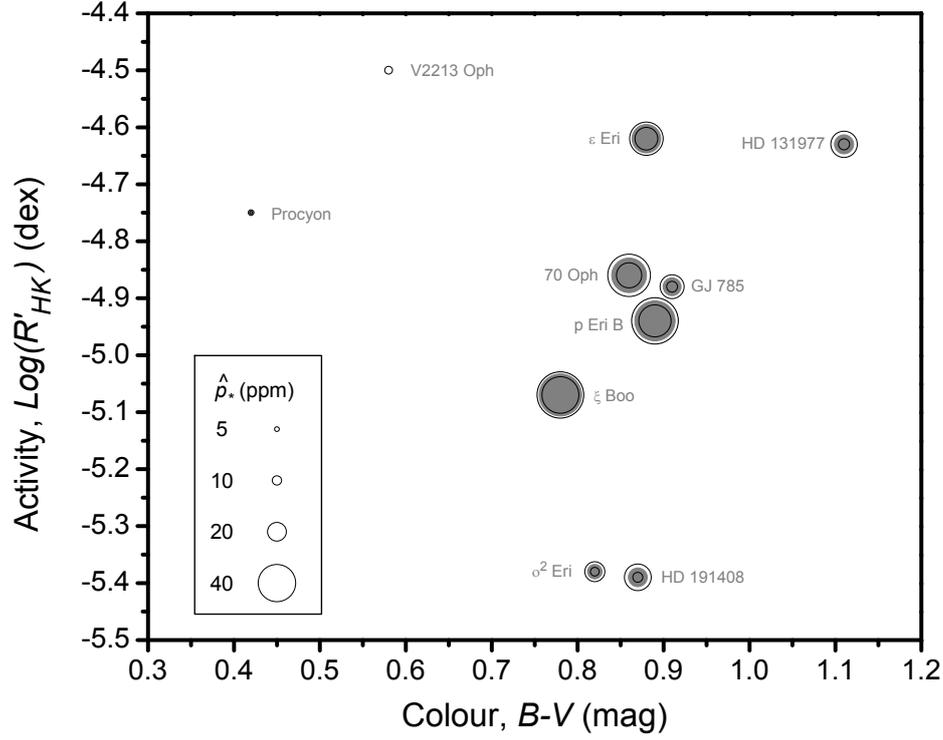}
\caption{A plot showing the determined intrinsic polarization of active stars in this study relative to their colour and activity index. The areas of the grey circles represent the debiased intrinsic polarization, whilst the solid circles are the 1-sigma errors.\label{fig:active}}
\end{figure*}

To test the qualitative predictions of \citet{saar93}'s models, in Figure \ref{fig:active} we have plotted the debiased intrinsic polarization of each active star against the activity indicator $log(R'_{HK})$ and the photometric colour ($B-V$)\footnote{For reference a nominal F0 dwarf has a characteristic temperature of 7200 K and a $B-V$ colour of 0.294; G0 5920 K and 0.588; K0 5280 K and 0.816; and K5 4450 K and 1.134 \citep{pecaut13}.}. This data is also tabulated in Table \ref{tab:active}. It should be noted that the literature values for $log(R'_{HK})$ come predominantly from \citet{martinez-arnaiz10}. Using a single literature source ensures consistency, but activity levels vary over time, and these measurements were more than 5 years old at the time of our observations. Bearing in mind this caveat, and those stated above, there is some support for the differential saturation models in the data. The two least polarized active stars, Procyon and V2213 Oph, are also the two with the lowest $B-V$ colour values. This is despite them being nominally more active.

There are seven active dwarfs with $B-V$ values between 0.75 and 0.90. Of those, two of the three least polarized have the lowest (most negative) activity indicies. The third is GJ 785, for which our $log(R'_{HK})$ value comes from \citet{jenkins06}. However, GJ 785 was also observed by \citet{martinez-arnaiz10} who classified it as not active. So, it should probably be much lower on the diagram. This then supports the notion that for a given temperature, less active stars are less polarized.

If the same global fields measured by spectropolarimetry with circular polarization are responsible for the broadband linear polarization we measure then we should see a correspondence between the net field and our measurements. To test this we have looked at data for stars we have in common with two spectropolarimetric surveys of magnetic fields, BCool \citep{marsden14} and PolarBase \citep{petit14}, and obtained determinations of the net longitudinal magnetic field ($B_l$), using the formulation of \citep{donati97}. For stars in the PolarBase database we downloaded the observations and created Stokes \textit{V} LSD (Least-Squares Deconvolution \citep{donati97}), using the same line masks used by the BCool collaboration \citep{marsden14}. To do this we assumed a stellar temperature for each star based on information in the PASTEL database \citep{soubiran16}. The velocity range over which $B_l$ is calculated has been chosen to maximise $|B_l|/\Delta B_l$ as in the BCool work \citep{marsden14}. We have tabulated minimum and maximum $B_l$ values obtained for stars we've classified as both active and inactive in Table \ref{tab:bl}.

\begin{table}
\caption{Longitudinal magnetic field measurements from BCool and PolarBase.}
\tabcolsep 2 pt
\centering
\begin{tabular}{lrcrlrr}
\hline
Name        & \multicolumn{1}{l}{HD}        &   $B-V$   &   Obs. & Ref.$^a$ &   \multicolumn{2}{c}{$|B_l|$ (G)}  \\
                    &           &           &        &          &   \multicolumn{1}{c}{max} &   \multicolumn{1}{c}{min}\\
\hline
\multicolumn{7}{l}{\textit{Active Stars}}  \\
$\epsilon$ Eri	    &   22049	&	0.88	&    58  &  B      &   -10.9~$\pm$~0.2            &   +0.4~$\pm$~0.2  \\
$\omicron^2$ Eri	&   26965	&	0.82	&     1  &  B      &   +1.3~$\pm$~0.2  &        \\
Procyon             &   61421	&	0.37	&     1  &  P      &   +2.0~$\pm$~0.7  &        \\
$\ksi$ Boo$^b$	    &   131156	&	0.78	&   101  &  B      &   +18.4~$\pm$~0.3            &   +0.5~$\pm$~1.0  \\
GJ 785          	&   192310	&	0.91	&     1  &  P      &   -3.9~$\pm$~0.2           \\
\hline
\multicolumn{7}{l}{\textit{Inactive Stars}}  \\
$\tau$ Cet          &   10700   &   0.57    &     2  &  P      &   -1.8~$\pm$~0.9  &   +1.2~$\pm$~0.8  \\
$\pi^3$ Ori         &   30652   &   0.45    &     5  &  B,P    &   -4.9~$\pm$~1.7  &   +1.3~$\pm$~0.7  \\
$\gamma$ Lep        &   38393   &   0.47    &     6  &  P      &   -3.3~$\pm$~1.3  &   -0.4~$\pm$~0.4  \\
$\beta$ Vir         &   102870  &   0.55    &     3  &  P      &   +3.3~$\pm$~0.9  &   -0.9~$\pm$~0.1  \\
61 Vir              &   115617  &   0.70    &     1  &  P      &   -0.2~$\pm$~0.2  &                   \\
\hline
\end{tabular}
\begin{flushleft}
a - P: PolarBase \citep{petit14}, B: BCool \citep{marsden14}. \\
b - The data presented are for HD 131156A; $|B_l|_{max}$ for HD 131156B is similar. \\
\end{flushleft}
\label{tab:bl}
\end{table}

\begin{figure*}
\centering
\includegraphics[width=0.75\textwidth,trim={0 0cm 0 0cm}]{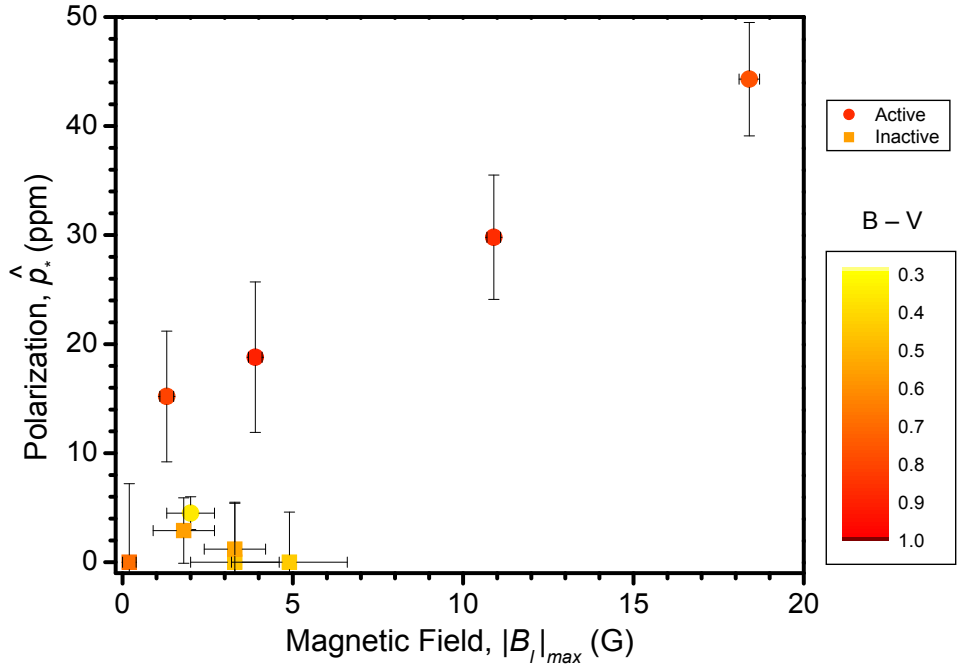}
\caption{A plot showing the determined intrinsic polarization of active stars in this study relative to their maximal longitudinal magnetic field determined from spectropolarimetric circular polarization measurements. The colour scale denotes the $B-V$ colour of the stars. Stars we classified as active in this study are denoted by circles, inactive stars by squares.\label{fig:bl}}
\end{figure*}

From Table \ref{tab:bl} there are many observations of $\epsilon$ Eri and $\ksi$ Boo, and we can be confident the strength of the field is captured by the observations. For a further three of the active stars there is only a single observation. The configuration of the magnetic field can vary substantially over/within a rotation period and an activity cycle -- large regions of positive and negative field can cancel each other out -- and as a result it is difficult to know if these measurements are representative. Similarly, we have a handful of observations for a further five stars we've classified as inactive.

We have plotted $\hat{p_{\star}}$ against the greatest field recorded for each star ($|B_l|_{max}$) in Figure \ref{fig:bl}. With the caveat that the data is sparse, the stars with the strongest longitudinal magnetic fields also recorded the greatest linear polarization. Although the stars with $|B_l|_{max}$ less than 5 G, have fewer observations, of those, the cooler stars ($B-V>0.75$) measured greater linear polarizations. This suggests that for cooler stars, which we expect to have greater line-blanketing, the magnetic field strength determined from circular polarization is predictive of linear polarization. Further work will need to be done to test this hypothesis, to determine how circular polarization is related to broadband linear polarization and how they are related over an activity cycle or modulated over a rotation period. It is also desirable to probe cooler objects like M-dwarfs where increased line-blanketing could potentially over-saturate the spectral lines within a band and reduce the measurable polarization. \citet{leroy90} concluded that the blending of spectral lines resulted in a limit to the maximum polarization from differential saturation, and that this was already reached in the blue part of the spectrum for solar types. The models of \citet{stift97} show polarization increasing with reducing temperature in the visible part of the spectrum, but at a decreasing rate on a per Kelvin basis.

\subsubsection{A comparison with previous linear polarization measurements of active dwarfs}

At this juncture it is pertinent to point out that the magnitudes of polarization we record in our data are well below those suggested by other studies of active late dwarfs. Most recently \citet{patel16} reported 800~$\pm$~60 ppm in V-band, and 1600~$\pm$~100 ppm in B-band. \citet{alekseev03} reports even higher levels of polarization than this. Whilst \citet{patel16} have observed, on average, more active stars, this seems in extraordinarily poor agreement with our mean in $g^{\rm \prime}$ of 23.0~$\pm$~2.2 ppm. The most polarized object in the work of \citet{huovelin88} is the same object that is most polarized of those we report here -- $\ksi$ Boo -- for it they report 400~$\pm$~60 ppm in V-band, and 340~$\pm$~140 ppm in B-band. The intrinsic polarization we calculate in $g^{\rm \prime}$ for $\ksi$ Boo is an order of magnitude less than this. Differential saturation is highly wavelength dependant, and the models of \citet{saar93} clearly indicate that polarization in B-band is expected to be a few times that in the $g^{\rm \prime}$ filter, but they also indicate that V-band polarization will be less than that of $g^{\rm \prime}$.

Whilst the contention of \citet{clarke91,leroy89} that \citet{huovelin88}'s measurements were affected by scattered moonlight has already been mentioned, \citet{patel16}'s results are new, and we need to examine why there might be a discrepancy between their results and ours. Their instrument made use of a rotatable half wave plate to provide polarizaiton modulation. Such a set-up is susceptible to scintillation noise due to seeing (see \citet{kemp87,hough06,wiktorowicz08,bailey15} and references within), and is not well suited to measuring polarization at the ppm level. This combined with the faintness of the targets they observed (all fainter than $m_V=$~6 due to limitations of their CCD detector) means it is probable that their observations are dominated by noise. It is difficult to know what to make of \citet{alekseev03}'s measurements given that he has made use of the same instrument, telescope and statistical techniques as \citet{huovelin88} but sees even larger polarizations. It is probable that these are also dominated by scintillation noise. \citet{alekseev03} attributed the difference between his measurements and the predictions of \citet{huovelin91,saar93} to additional circumstellar material; something we don't see evidence for in our results for ordinary FGK dwarfs.

\citet{kochukhov11} used the HARPSpol instrument \citep{snik11,piskunov11} installed at the Cassegrain focus of the 3.6 m telescope at the European Southern Observatory to make observations of three stars including $\epsilon$ Eri in 2011. HARPSpol is a spectropolarimeter (operating over a wavelength range of 380 to 690 nm) that makes use of Least Squares Deconvolution (LSD) to determine a polarization. \citet{kochukhov11} monitored $\epsilon$ Eri for the 11 nights of its rotation period, and report linear polarization determinations from the LSD amplitudes in $q$ and $u$ for the three nights where they have sufficient signal to noise for marginal or definite detections. Their results are reported in terms of individual Stokes parameters for the 8th, 9th and 11th of January, where they say the measurements are significant. The values they report for $q$ and $u$ are of the same order as those we report here, but are not directly comparable. The equivalent width obtained from an LSD profile can be equated to a broadband linear polarization measurement if one knows what scaling factor needs to be applied \citep{wade00,silvester12}. In Ap stars this scaling factor is of order $\sim$3 to 10 and is always applied to increase the broadband measurement to account for the fact that many regions of the band contain no significant spectral lines. In cooler dwarfs, with a greater number of spectral lines we might expect it to be less. However, the relationship between the LSD amplitude (reported by \citet{kochukhov11}) and the equivalent width is not readily predictable \citep{wade00,silvester12}, meaning we cannot compare our results with \citet{kochukhov11}'s to determine a scaling factor.

The observations of $\epsilon$ Eri use similar sized telescopes with both instruments at the Cassegrain focus, which affords us the opportunity to compare the relative sensitivities of aperture and LSD techniques. Our total exposure time was 640~s, and we achieved a formal error of 5.3~ppm. \citet{kochukhov11}'s best precision is reported for the 8th of January. On that night their exposure time was 2000~s. Meaning that a roughly 8000~s exposure would be necessary to achieve the same precision as our broadband aperture technique assuming photon-limited performance.

\subsubsection{Implications for exoplanet polarimetry}

The hot Jupiter system HD 189733 is the most promising target for the detection of scattered light from a planetary atmosphere. HD 189733b has been the subject of an ongoing controversy in the literature, a succinct run-down of which can be found in the recent work of \citet{bott16}. In short, whilst observations by \citet{berdyugina08,berdyugina11} report polarization levels of $\sim$~100 ppm, where the largest signals are at the shortest wavelengths, these results have not been replicated by \citet{wiktorowicz09,wiktorowicz15b}. Nor by \citet{bott16} who reports a possible polarization amplitude, matching the phase of the planet, of 29.4~$\pm$~15.6 ppm (in a 500 nm short pass band which had an effective wavelength of 446.1 nm).

HD 189733 is a K1.5V ($B-V=0.93$) BY Dra variable with starspot coverage $\sim$~2 per cent, and from a determination we made from 96 observations in the PolarBase database \citep{petit14}, an extreme value of ${B_l}$ of -17.3~$\pm$~0.7 G. Calculations of the effect of starspots on HD 189733 based on their partial occulting of the stellar disc by \citet{kostogryz15} give only a 3 ppm variation. The possible effects of differential saturation modulated by stellar activity are mentioned by \citet{bott16} but appear not to have been considered by the other authors and other stellar effects are usually assumed to be negligible. Our results presented here reveal linear polarization of 10s of ppm for BY Dra variables with similar spectral types to HD 189733, which may be attributable to differential saturation associated with similarly strong magnetic fields. If differential saturation from starspots causes polarization in these stars, then it may also help explain the different results seen for HD 189733 given that stellar activity can be variable on timescales of years. Although planet's orbital period ($\sim$2.2 d \citep{triaud10}) and the stellar rotation period ($\sim$11.8 d \citep{moutou07}) are quite different, the precision of the measurements at present, and the fact that \citet{berdyugina08,berdyugina11} have not reported the exact timings of their measurements, mean that it is not possible to determine whether stellar activity has been mistaken for a planetary signal through signal aliasing.

The baseline polarization (i.e. that not tentatively attributed to the planet) determined by \citet{bott16} for HD 189733 was 70.1~$\pm$~9.6 ppm, at a polarization angle of 20.0~$\pm$~3.9$\degr$, which they note is larger than is expected from interstellar polarization alone. Our model gives $p_i$ as 20.0 ppm after a minor adjustment for the bluer effective wavelength. There are only three stars in the Interstellar List within 35$\degr$ of HD 189733, and the closest has quite a large uncertainty, and doesn't agree well with the other two; if we cautiously remove it, we make $\theta_i$ to be 14.4$\degr$. Subtracting the predicted interstellar polarization from the baseline gives $\hat{p}=$~52.6~$\pm$~9.6 ppm as the likely contribution from stellar activity. This is similar to, but a little higher than determined for $\ksi$ Boo which is a warmer star but has a similar $|B_l|_{max}$ value. So, the determination is consistent with differential saturation producing higher polarization at bluer wavelengths and/or in cooler stars with similar activity levels.

The nature of stellar activity may be similarly important for observations of $\tau$ Boo. A signal from $\tau$ Boo b was unsuccessfully sought by \citet{lucas09}, who, however, noted that greater scatter in their results correlated with starspot activity. In that system a large hotspot has been observed leading the subplanetary point by 60$\degr$ to 70$\degr$, and starspots have also been seen to move in phase with the planet \citep{lanza08}. $\tau$ Boo is a warmer star ($B-V=0.49$) than those we recorded significant polarizations for here, and its extreme value of ${B_l}$ from 177 observations from BCool and PolarBase is +4.6~$\pm$~0.4 G. These facts combined suggest that the contribution of magnetic activity is likely to be smaller than for HD 189733.

Because the rotation period of HD 189733 is different from the orbital period of HD 189733b, a starspot signal could be removed. Performing a similar task may be more difficult for $\tau$ Boo b but with better characterisation of polarization due to differential saturation it should be possible. \citet{saar93} and \citet{stift97}'s models predict signals from differential saturation that are highly wavelength dependant and do not vary smoothly. In contrast the polarization due to Rayleigh scattering from exoplanetary clouds can be expected to be a fairly smooth function of wavelength which rises to the blue \citep{evans13}. Therefore simultaneous observations in multiple pass bands could be used to decouple the two effects.

\section{Conclusions}
\label{sec:conclusions}

We have made a short linear polarimetric survey of nearby FGK dwarfs at ppm precision. Amongst the sample were debris disk host stars and active stars. Our initial statistical analysis showed active stars to be more polarized than ordinary FGK dwarfs, but no discernable difference between inactive debris disk stars and other inactive stars.

We added our data on ordinary FGK dwarfs to literature measurements of other nearby stars which improved our knowledge of the local ISM. The data shows some alignment with the local ISM field, but we don't see the same Loop I Superbubble structure associated with the region at 50 to 100 pc. We find that there are broadly two regions with differing polarization with distance relations within 100 pc. Although the ISM is patchy, above $b=+30$ we make the relationship to be: \begin{equation}p_i=(0.261\pm0.017)d,\end{equation} where $p_i$ is in ppm and $d$ is in pc. For the stars below $b=+30$ and within 14.5 pc we find: \begin{equation} p_i=(0.800~\pm~0.120)d,\end{equation} below $b=+30$ and beyond 14.5 pc the relationship is:\begin{equation}p_i=(1.644~\pm~0.298)(d-14.5) + (11.6~\pm~1.7).\end{equation} We also determined that the position angles of stars polarized by the ISM are increasingly well aligned for separations decreasing from 35$\degr$.

We used the information obtained on the ISM to construct a simple model for determining interstellar polarization. Up until now subtractions of interstellar polarization within the LHB have required additional measurements of control stars. Our model will become increasingly powerful as more measurements of nearby stars are added at ppm precisions, potentially eliminating the need for control measurements. This development will drastically reduce the time required for precision polarimetric work on nearby objects.

After subtracting interstellar polarization using our model we find the mean polarization of the active stars to be 23.0~$\pm$~2.2 ppm compared to 2.9~$\pm$~1.9 ppm for the inactive non-debris disk stars. The most polarized star in our survey was the active star $\ksi$ Boo at 44.3~$\pm$~5.2 ppm. Both these figures are much less than reported by other researchers. Although the data may be explained by polarization either through Rayleigh scattering from large ($\sim$~18 per cent) starspots at close to optimal alignments, or differential saturation from localised regions of strong magnetic fields like starspots with filling factors greater than $\sim$~0.25 to 2.0 per cent, we suggest differential saturation attributable to weaker global scale magnetic fields to be the most likely mechanism. The most polarized active stars also have large net longitudinal magnetic fields. The result has important implications for efforts to detect scattered polarized light from hot Jupiter clouds in the combined light of star and planet. Positive detections for planets orbiting active stars will be more challenging than previously assumed. An improved understanding of intrinsic polarization in active stars will help overcome the challenges.

For debris disk host stars we find a mean of 7.8~$\pm$~2.9 ppm after interstellar subtraction, with a marginal 3-sigma detection for the disk around HD 207129 which amounts to a polarization $\sim$~30 per cent of its infrared excess. Upon examining our data conscientiously system by system we can explain the signals in most systems by the disk infrared excess and geometry with respect to the aperture. A high $p:(L_{dust}/L_{\star})$ ratio for $\zeta^2$ Ret corroborates literature reports that its disk is asymmetric.

\section*{Acknowledgments}

This work was supported by the Australian Research Council through Discovery Project grant DP160103231. JPM is supported by a UNSW VC's Postdoctoral Research Fellowship. The authors thank the Director and staff of the Australian Astronomical Observatory for their advice and support with interfacing HIPPI to the AAT and during our observing runs on the telescope. We thank Daniela Optiz and Gesa Gruening for their assistance in making observations in February/March and June 2016 respectively. We acknowledge the use of the SIMBAD database and this research has made use of the VizieR Service at Centre de Donn\'{e}es Astronomiques de Strasbourg. We wish to thank the anonymous referee for valuable feedback that has improved this paper.

\bibliographystyle{mnras}
\bibliography{hr_study}

\appendix

\section{Stars Representative of Interstellar Polarization}
\label{apx:interstellar}

Table \ref{tab:interstellar_list} lists 58 stars observed with either HIPPI or PlanetPol which are believed to have polarization characteristic of the interstellar medium. The 14 inactive non-debris disk FGK dwarfs and e Eri and $\eta$ Crv observations reported here are used in conjunction with these tabulated observations to define interstellar polarization in this paper. In the table all objects cited as \citet{bailey10} have been converted to $g^{\rm \prime}$ by multiplication of the magnitude of polarization by 1.2, see \citet{marshall16}. Slight differences between the other results tabulated here and the original reference are attributable to minor improvements made in the post-observation analysis pipeline since publication. Polarization angle errors quoted here are derived in the same way as those in the main body of the paper, regardless of how they were reported in the original reference. Low polarization standards where the updated measurements reported correspond to the aggregate measurements of a number of observing runs.

Some stars have been excluded from some or all parts of the interstellar determinations made in the paper as follows: 
\begin{enumerate}
\item HD 7693 is excluded from Section \ref{sec:p_vs_d} as an outlier in polarization magnitude, but included in determining the angle interstellar polarization in later sections -- the angle determined for its primary HD 7788 is very similar, see \citet{marshall16}. 
\item HD 28556 is excluded from determinations of interstellar polarization magnitude and angle due to the size of the error, and the relatively few stars at that distance. 
\item $\alpha$ Hya and $\epsilon$ Dra are included in the $p/d$ determination for the $b>+30$ group rather than the $b<+30$ group. 
\item Arcturus exhibits variable polarization in the B-band according to \citet{kemp83}, the measurement utilised here corresponds to the red (575-1025 nm) bandpass of PlanetPol. 
\item $\alpha^2$ Lib is excluded from analysis on account of a discrepant magnitude and angle of polarization relative to neighbouring stars in a region where both are well defined.  
\item Stars at greater than 50 pc are not used in Sections \ref{sec:interstellar_angle} and \ref{sec:isub}.
\end{enumerate}

\begin{table*}
\caption{Additional literature stars for the Interstellar List}
\tabcolsep 3 pt
\centering
\begin{tabular}{rlccrrrrrrll}
\hline
HD & Sp. Type & \multicolumn{1}{c}{RA} & \multicolumn{1}{c}{Dec} &  \multicolumn{2}{c}{Galactic} & \multicolumn{1}{c}{$d$} & \multicolumn{1}{c}{$p$$^a$} & \multicolumn{1}{c}{$\theta$} & \multicolumn{1}{c}{$\theta_G$} &   Ref &   Notes$^b$ \\   
        &                  &  (hh mm ss.s)     &   (dd mm ss)      & $l$ ($\degr$)   & $b$ ($\degr$)   & \multicolumn{1}{c}{(pc)} & \multicolumn{1}{c}{(ppm)} & \multicolumn{1}{c}{($\degr$)} & \multicolumn{1}{c}{($\degr$)} & &  \\
\hline
739     &   F5V            &   00 11 44.0      &   -35 07 59       &   347.18      &   -78.34      &   21.3    &   39.3~$\pm$~11.4           &   71.0~$\pm$~\phantom{0}8.6 &   120.2           &   \citet{marshall16}  &           \\
2151    &   G0V            &   00 25 39.2      &   -77 15 18       &   304.78      &   -39.78      &    7.5    &    5.8~$\pm$~\phantom{0}1.4 &   95.4~$\pm$~\phantom{0}7.2 &   102.6           &   Updated.            &   L       \\
2261    &   K0.5IIIb       &   00 26 16.9      &   -42 18 18       &   320.00      &   -73.97      &   23.7    &   18.9~$\pm$~\phantom{0}5.5 &   62.3~$\pm$~\phantom{0}8.6 &    82.8           &   \citet{cotton16b}   &           \\
4128    &   K0III          &   00 43 35.2      &   -17 59 12       &   111.34      &   -80.68      &   29.4    &   23.5~$\pm$~\phantom{0}6.6 &   95.1~$\pm$~\phantom{0}8.3 &   106.0           &   \citet{cotton16}    &           \\
4308    &   G8V            &   00 44 39.0      &   -65 38 52       &   304.06      &   -51.46      &   21.9    &   24.6~$\pm$~15.5           &  120.4~$\pm$~22.5           &   122.7           &   \citet{marshall16}  &           \\
7693    &   K2V+K3V        &   01 15 01.0      &   -68 49 08       &   299.75      &   -48.16      &   21.6    &  162.8~$\pm$~22.7           &   96.8~$\pm$~\phantom{0}4.0 &    88.8           &   \citet{marshall16}  &   i       \\
12311   &   F0IV           &   01 58 45.9      &   -61 34 12       &   289.45      &   -53.76      &   21.9    &   42.4~$\pm$~\phantom{0}7.2 &  159.2~$\pm$~\phantom{0}4.9 &   133.3           &   \citet{marshall16}  &           \\
18622J  &   A4III          &   02 58 15.7      &   -40 18 17       &   247.84      &   -60.73      &   49.5    &   74.0~$\pm$~\phantom{0}7.1 &   31.5~$\pm$~\phantom{0}2.8 &   138.7           &   \citet{cotton16}    &           \\
28556   &   F0IV           &   04 30 37.4      &    13 43 28       &   182.50      &   -22.97      &   45.2    &   50.8~$\pm$~26.1           &    7.1~$\pm$~18.0           &   135.2           &   \citet{marshall16}  &   ii      \\
45348   &   A9II           &   06 23 57.1      &   -52 41 45       &   261.22      &   -25.29      &   95.9    &  113.0~$\pm$~\phantom{0}1.2 &  116.2~$\pm$~\phantom{0}0.3 &    38.7           &   \citet{bailey15}    &   v       \\
48915   &   A0V            &   06 45 09.2      &   -16 42 47       &   227.23      &    -8.89      &    2.6    &    1.8~$\pm$~\phantom{0}0.6 &  163.3~$\pm$~11.4           &    99.3           &   Updated.            &   L       \\
74956   &   A1Va(n)        &   08 44 42.2      &   -54 42 32       &   272.08      &    -7.37      &   24.7    &   44.5~$\pm$~\phantom{0}6.8 &  127.9~$\pm$~\phantom{0}4.4 &    75.7           &   \citet{cotton16}    &           \\
80007   &   A1III          &   09 13 12.0      &   -69 43 02       &   285.98      &   -14.41      &   34.7    &   23.9~$\pm$~\phantom{0}2.6 &   74.4~$\pm$~\phantom{0}3.1 &    25.8           &   \citet{cotton16}    &           \\
81797   &   K3II-III       &   09 27 35.2      &   -08 39 31       &   241.49      &    29.05      &   55.3    &    8.8~$\pm$~\phantom{0}3.7 &  170.8~$\pm$~13.7           &   118.7           &   \citet{cotton16}    &   iii, v \\
89025   &   F0IIIa         &   10 16 41.4      &    23 25 01       &   210.22      &    54.95      &   84.0    &   16.9~$\pm$~\phantom{0}2.9 &  107.3~$\pm$~\phantom{0}5.0 &    32.2           &   \citet{bailey10}    &   v       \\
93497   &   G6III          &   10 46 46.1      &   -49 25 12       &   283.03      &     8.57      &   35.5    &   33.5~$\pm$~\phantom{0}4.3 &  123.4~$\pm$~\phantom{0}3.7 &    95.6           &   \citet{cotton16}    &           \\
95689   &   G9III+A7V      &   11 03 43.7      &    61 45 04       &   142.85      &    51.01      &   37.7    &   11.2~$\pm$~\phantom{0}1.1 &  141.5~$\pm$~\phantom{0}2.8 &   101.9           &   \citet{bailey10}    &           \\
96833   &   K1III          &   11 09 39.8      &    44 29 55       &   165.80      &    63.23      &   44.3    &    4.6~$\pm$~\phantom{0}2.5 &  109.3~$\pm$~19.3           &    51.6           &   \citet{bailey10}    &           \\
97603   &   A5IV(n)        &   11 14 06.5      &    20 31 25       &   224.24      &    66.83      &   17.9    &    4.4~$\pm$~\phantom{0}2.8 &  158.6~$\pm$~22.7           &    90.2           &   \citet{bailey10}    &           \\
97633   &   A2IV           &   11 14 14.4      &    15 25 47       &   235.37      &    64.59      &   50.6    &    8.3~$\pm$~\phantom{0}3.2 &   25.2~$\pm$~12.6           &   146.8           &   \citet{bailey10}    &   v       \\
102224  &   K0.5IIIb       &   11 46 03.0      &    47 46 46       &   150.32      &    65.72      &   56.3    &   12.1~$\pm$~\phantom{0}3.1 &  148.9~$\pm$~\phantom{0}7.6 &   111.6           &   \citet{bailey10}    &   v       \\
102509  &   A7V            &   11 48 59.1      &    20 13 08       &   235.01      &    73.94      &   71.3    &   14.0~$\pm$~\phantom{0}4.4 &  100.5~$\pm$~\phantom{0}9.4 &    39.2           &   \citet{bailey10}    &   v       \\
108767  &   A0IV           &   12 29 51.9      &   -16 30 57       &   295.48      &    46.04      &   26.6    &   18.7~$\pm$~\phantom{0}3.8 &   70.0~$\pm$~\phantom{0}5.8 &    63.0           &   \citet{cotton16}    &           \\
109379  &   G5II           &   12 34 24.0      &   -23 23 48       &   297.88      &    39.30      &   42.8    &   34.5~$\pm$~\phantom{0}3.8 &   76.1~$\pm$~\phantom{0}3.2 &    71.1           &   \citet{cotton16}    &           \\
110304  &   A1IV+A0IV      &   12 41 31.0      &   -48 57 36       &   301.26      &    13.88      &   39.9    &   61.6~$\pm$~\phantom{0}4.2 &   63.8~$\pm$~\phantom{0}1.9 &    61.4           &   \citet{cotton16}    &           \\
110379J &   F0IV+F0IV      &   12 41 39.6      &   -01 26 58       &   297.84      &    61.33      &   11.7    &    7.6~$\pm$~\phantom{0}4.0 &  112.5~$\pm$~18.6           &   107.9           &   \citet{cotton16}    &           \\
113226  &   G8III          &   13 02 10.6      &    10 57 33       &   312.31      &    73.64      &   33.6    &    6.8~$\pm$~\phantom{0}1.2 &    3.1~$\pm$~\phantom{0}5.1 &    11.5           &   \citet{bailey10}    &           \\
115659  &   G8III          &   13 18 55.2      &   -23 10 17       &   311.10      &    39.27      &   40.5    &    5.6~$\pm$~\phantom{0}4.2 &   36.4~$\pm$~26.2           &    44.2           &   \citet{cotton16}    &           \\
116656  &   A1.5Vas        &   13 23 55.5      &    54 55 31       &   113.11      &    61.57      &   26.3    &    9.1~$\pm$~\phantom{0}1.9 &  173.2~$\pm$~\phantom{0}5.9 &     8.6           &   \citet{bailey10}    &           \\
121370  &   G0IV           &   13 54 41.1      &    18 23 52       &     5.29      &    73.03      &   11.4    &    4.2~$\pm$~\phantom{0}2.0 &  167.4~$\pm$~16.0           &    43.3           &   \citet{bailey10}    &           \\
123129  &   K0III          &   14 06 40.9      &   -36 22 12       &   319.45      &    24.08      &   18.0    &   42.6~$\pm$~\phantom{0}3.3 &   56.7~$\pm$~\phantom{0}2.2 &    74.9           &   \citet{cotton16}    &           \\
124897  &   K0III          &   14 15 39.7      &    19 10 57       &    15.05      &    69.11      &   11.3    &    7.5~$\pm$~\phantom{0}1.8 &   30.7~$\pm$~\phantom{0}7.0 &    94.1           &   \citet{bailey10}    &   iv      \\
127665  &   K3III          &   14 31 49.8      &    30 22 17       &    47.29      &    67.80      &   49.1    &   12.3~$\pm$~\phantom{0}2.8 &  124.5~$\pm$~\phantom{0}6.7 &    30.2           &   \citet{bailey10}    &           \\
127762  &   A7IV(n)        &   14 32 04.7      &    38 18 30       &    67.27      &    66.17      &   26.6    &    4.3~$\pm$~\phantom{0}1.9 &  109.1~$\pm$~15.2           &   178.3           &   \citet{bailey10}    &           \\
130841  &   A3V            &   14 50 52.8      &   -16 02 30       &   340.33      &    38.01      &   23.2    &   27.8~$\pm$~\phantom{0}3.4 &  142.1~$\pm$~\phantom{0}3.4 &   176.2           &   \citet{cotton16}    &   v       \\
140573  &   K2IIIb         &   15 44 16.1      &    06 25 32       &    14.21      &    44.08      &   22.7    &    4.4~$\pm$~\phantom{0}1.7 &   16.3~$\pm$~12.3           &    74.1           &   Updated.            &   L       \\
147547  &   A9IIIbn        &   16 21 55.2      &    19 09 11       &    35.25      &    41.30      &   59.1    &   15.9~$\pm$~\phantom{0}3.2 &   47.8~$\pm$~\phantom{0}5.9 &   117.7           &   \citet{bailey10}    &   v       \\
148856  &   G7IIIa~Fe      &   16 30 13.2      &    21 29 23       &    39.01      &    40.21      &   42.7    &   22.5~$\pm$~\phantom{0}1.6 &   24.5~$\pm$~\phantom{0}2.1 &    96.1           &   \citet{bailey10}    &           \\
150680  &   G0IV           &   16 41 17.2      &    31 36 10       &    52.66      &    40.29      &   10.7    &   11.5~$\pm$~\phantom{0}3.1 &   51.7~$\pm$~\phantom{0}8.0 &   130.7           &   \citet{bailey10}    &           \\
151680  &   K1III          &   16 50 10.2      &   -34 17 33       &   348.81      &     6.56      &   20.1    &   28.9~$\pm$~\phantom{0}6.7 &   34.4~$\pm$~\phantom{0}6.8 &    84.8           &   \citet{cotton16}    &           \\
153210  &   K2III          &   16 57 40.1      &    09 22 30       &    28.37      &    29.50      &   28.0    &   14.4~$\pm$~\phantom{0}1.9 &   64.1~$\pm$~\phantom{0}3.8 &   127.9           &   \citet{bailey10}    &           \\
153808  &   A0V            &   17 00 17.4      &    30 55 35       &    52.86      &    36.18      &   47.5    &   14.9~$\pm$~\phantom{0}4.7 &   66.0~$\pm$~\phantom{0}9.5 &   142.8           &   \citet{bailey10}    &           \\
155125  &   A2IV-V         &   17 10 22.7      &   -15 43 30       &     6.71      &    14.01      &   25.8    &   57.2~$\pm$~\phantom{0}3.6 &  147.7~$\pm$~\phantom{0}1.8 &    23.5           &   \citet{cotton16}    &           \\
156164  &   A1IVn+G4IV-V   &   17 15 01.9      &    24 50 21       &    46.83      &    31.42      &   23.0    &    9.4~$\pm$~\phantom{0}2.9 &   66.2~$\pm$~\phantom{0}9.1 &   138.0           &   \citet{bailey10}    &           \\
159532  &   F1III          &   17 37 19.1      &   -43 59 52       &   347.14      &    -5.98      &   83.4    &  154.4~$\pm$~\phantom{0}2.9 &   94.0~$\pm$~\phantom{0}0.6 &   151.7           &   \citet{cotton16}    &   v       \\
159561  &   A5III          &   17 34 56.1      &    12 33 36       &    35.89      &    22.57      &   14.9    &   28.1~$\pm$~\phantom{0}2.4 &   30.8~$\pm$~\phantom{0}2.4 &    96.1           &   \citet{bailey10}    &           \\
161096  &   K2III          &   17 43 28.4      &    04 34 02       &    29.22      &    17.19      &   25.1    &   38.6~$\pm$~\phantom{0}2.6 &   27.9~$\pm$~\phantom{0}1.9 &    90.6           &   \citet{bailey10}    &           \\
161797  &   G5IV           &   17 46 27.5      &    27 43 14       &    52.44      &    25.63      &    8.3    &   11.1~$\pm$~\phantom{0}2.5 &   21.0~$\pm$~\phantom{0}6.4 &    92.0           &   \citet{bailey10}    &           \\
163588  &   K2III          &   17 53 31.7      &    56 52 22       &    85.16      &    30.23      &   34.5    &    4.5~$\pm$~\phantom{0}3.7 &   51.1~$\pm$~28.1           &   136.3           &   \citet{bailey10}    &           \\
163993  &   G8III          &   17 57 45.9      &    29 14 52       &    54.91      &    23.77      &   41.9    &   29.2~$\pm$~\phantom{0}2.9 &   10.1~$\pm$~\phantom{0}2.8 &    80.8           &   \citet{bailey10}    &           \\
164058  &   K5III          &   17 56 36.4      &    51 29 20       &    79.06      &    29.22      &   47.3    &   87.9~$\pm$~\phantom{0}1.4 &  145.0~$\pm$~\phantom{0}0.5 &    46.7           &   \citet{bailey10}    &           \\
165135  &   K1III          &   18 05 48.5      &   -30 25 25       &     0.92      &    -4.53      &   29.5    &   38.5~$\pm$~16.1           &  173.8~$\pm$~13.8           &    54.5           &   \citet{cotton16}    &           \\
168775  &   K2III          &   18 19 51.7      &    36 03 52       &    63.52      &    21.54      &   77.2    &  127.8~$\pm$~\phantom{0}3.2 &    6.1~$\pm$~\phantom{0}0.7 &    77.2           &   \citet{bailey10}    &   v       \\
169916  &   K0IV           &   18 27 58.3      &   -25 25 16       &     7.66      &    -6.52      &   20.7    &   54.2~$\pm$~\phantom{0}8.2 &  140.1~$\pm$~\phantom{0}4.5 &    22.8           &   \citet{cotton16}    &           \\
176687  &   A2.5Va+A4IV    &   19 02 36.7      &   -29 52 48       &     6.84      &   -15.35      &   27.3    &   28.2~$\pm$~\phantom{0}3.6 &  135.9~$\pm$~\phantom{0}3.7 &    22.7           &   \citet{marshall16}  &           \\
187691  &   F8V            &   19 51 01.6      &    10 25 57       &    49.14      &    -8.20      &   19.2    &   18.9~$\pm$~\phantom{0}9.7 &  119.3~$\pm$~17.8           &   179.4           &   \citet{marshall16}  &           \\
188119  &   G8III+F5III    &   19 48 10.4      &    70 16 17       &   102.43      &    20.83      &   45.4    &    1.8~$\pm$~\phantom{0}3.1 &   63.4~$\pm$~40.2           &   130.9           &   \citet{bailey10}    &   iii     \\
207098  &   A7III          &   21 47 02.3      &   -16 07 48       &    37.60      &   -46.01      &   11.8    &   29.6~$\pm$~16.2           &  137.3~$\pm$~19.4           &    24.4           &   \citet{cotton16}    &           \\
\hline
\end{tabular}
\begin{flushleft}
a - $g^{\rm \prime}$ band equivalent. \\
b - L indicates a low polarization standard. For numeric notes see the text of Appendix \ref{apx:interstellar}. \\
\end{flushleft}
\label{tab:interstellar_list}
\end{table*}

\bsp
\label{lastpage}

\end{document}